\documentclass[final,onefignum,onetabnum]{siamonline220329} 

\usepackage{lipsum}
\usepackage{amsfonts}
\usepackage{amsmath}
\usepackage{amssymb}
\usepackage{graphicx}
\usepackage{epstopdf}
\usepackage{algorithmic}
\usepackage{mathrsfs}
\usepackage{caption}
\usepackage{subcaption}
\usepackage{blindtext}
\usepackage{pdflscape}
\ifpdf
  \DeclareGraphicsExtensions{.eps,.pdf,.png,.jpg}
\else
  \DeclareGraphicsExtensions{.eps}
\fi

\newcommand{\bra}[1]{\left(#1\right)}
\newcommand{\Bra}[1]{\left[#1\right]}
\newcommand{\BRA}[1]{\left\{#1\right\}}

\def\mathbi#1{\textbf{\em #1}}

\usepackage{enumitem}
\setlist[enumerate]{leftmargin=.5in}
\setlist[itemize]{leftmargin=.5in}

\newcommand{\norm}[1]{\left\lVert#1\right\rVert}

\newsiamremark{remark}{Remark}
\newsiamremark{hypothesis}{Hypothesis}
\crefname{hypothesis}{Hypothesis}{Hypotheses}
\newsiamthm{claim}{Claim}

\headers{Bistability of waves and wave-pinning states 
}{Hughes, Modai, Edelstein-Keshet, and Yochelis}

\title{Bistability of travelling waves and wave-pinning states in a mass-conserved reaction-diffusion system:\\From bifurcations to implications for actin waves
\thanks{Submitted to the editors \today.
\funding{This work was funded by a Natural Sciences and Engineering Research Council of Canada (NSERC) CGS-D Scholarship awarded to JMH, by an NSERC Discovery Grant to LEK., and by the United States - Israel Binational Science Foundation (BSF, grant no. 2022072), Jerusalem, Israel to AY.}}}

\author{Jack M. Hughes\thanks{Department of Mathematics, University of British Columbia, Vancouver, Canada 
  (\email{jhughes@math.ubc.ca}, \email{keshet@math.ubc.ca}).}
\and Saar O. Modai\thanks{Department of Physics, Ben-Gurion University of the Negev, Be’er Sheva 8410501, Israel
  (\email{modais@post.bgu.ac.il}).}
\and Leah Edelstein-Keshet\footnotemark[2]
\and Arik Yochelis\thanks{Swiss Institute for Dryland Environmental and Energy Research, Blaustein Institutes for Desert Research, Ben-Gurion University of the Negev, Sede Boqer Campus, Midreshet Ben-Gurion 8499000, and Department of Physics, Ben-Gurion University of the Negev, Be'er Sheva 8410501, Israel (\email{yochelis@bgu.ac.il}).}}

\usepackage{amsopn}

\begin{document}

\maketitle

\begin{abstract}
Eukaryotic cells demonstrate a wide variety of dynamic patterns of filamentous actin (F-actin) and its regulators. Some of these patterns play important roles in cell functions, such as distinct motility modes, which motivate this study. We devise a mass-conserved reaction-diffusion model for active and inactive Rho-GTPase and F-actin in the cell cortex. The mass-conserved Rho-GTPase system promotes F-actin, which feeds back to inactivate the former. We study the model on a 1D periodic domain (edge of thin sheet-like cell) using bifurcation theory in the framework of spatial dynamics, complemented with numerical simulations. Among several discussed bifurcations, the analysis centers on the study of the codimension-2 long wavelength and finite wavenumber Hopf instability, in which we describe a rich structure of steady wave-pinning states (a.k.a.\ mesas, obeying the Maxwell construction), propagating coherent solutions (fronts and excitable pulses), and travelling and standing waves, all distinguished by mass conservation regimes and classified by domain sizes. Specifically, we highlight the unexpected conditions for bistability between steady wave-pinning and travelling wave states on moderate domain sizes{, i.e., unfolding through domain length}. These results uncover and exemplify possible mechanisms of coexistence, robustness, and transitions between distinct cellular motility modes, including directed migration, turning, and ruffling. More broadly, the results indicate that non-gradient reaction-diffusion models comprising mass conservation have distinct pattern formation mechanisms that motivate further investigations, such as the unfolding of codimension-3 instabilities and T-points.
\end{abstract}



\section{Introduction}

In his work on morphogenesis~\cite{turing1952}, Alan Turing was concerned with spontaneous symmetry-breaking possibly leading to patterns of chemicals (``morphogens'') governing cell fates in embryo development (``morphogenesis’’). Since that time, exotic patterns of various proteins have even been visualized within eukaryotic cells~\cite{bailles2022mechanochemical,Bement2015,bernitt2017fronts}, and associated with cellular functions such as division or crawling. In the context of cell motility, patterns of proteins (``Rho-GTPases'') that regulate the assembly and distribution of filamentous actin (F-actin, a biopolymer) play an important role. Rho-GTPases can spontaneously self-organize, creating ``chemical prepatterns'' that define the ``front'' (Rac) or ``rear'' (Rho) of a cell (a.k.a.\ cell polarization)~\cite{goryachev2017many}. At the front, the GTPase Rac promotes F-actin assembly, which powers the protrusion of the cell edge and can induce cell motility. These actin dynamics are important for directed migration of eukaryotic cells, such as white blood cells that move toward sites of inflammation~\cite{weiner1999spatial,wong2006neutrophil}. It also governs the formation of the equatorial division furrow in a dividing cell \cite{Goryachev2016}. Mutations of the Rho-GTPase pathways can also contribute to the occurrence of metastatic cancer~\cite{xiao2021macropinocytosis}.

While F-actin assembly is typically downstream of Rho-GTPases (such as Rac or Rho), in some cases, there is also evidence for negative feedback from F-actin to the GTPase (e.g. by recruiting a ``GAP'', an effector that causes inactivation of the GTPase). Broadly, the dual interactions between GTPases and F-actin result in a variety of ``exotic'' dynamic patterns collectively known as ``actin waves''~\cite{Arai2010,barnhart2017adhesion,Bement2015,bernitt2017fronts,beta2017intracellular,dobereiner2006lateral,gerisch2009self,giannone2004periodic,Landino2021,Michaud2021,vicker_f_actin_2002}. Among the experimental observations, we find rhythmic waves of F-actin and protrusion of the cell edge (Fig.\ 1 in~\cite{giannone2004periodic}), lateral waves along a cell edge (Fig.\ 1 in~\cite{barnhart2017adhesion} and space-time plots in~\cite{dobereiner2006lateral}), interconversion of travelling and standing waves along a cell diameter (Fig.\ 1 in~\cite{wu2013calcium}) and waves on the cell cortex (Fig.\ 1 in~\cite{yang2018rhythmicity}). The presence of these dynamic events appears to be ubiquitous, while the proposed underlying mechanisms vary. A rich collection of experimental, theoretical, and joint experiment-modelling literature is dedicated to exploring such phenomena~\cite{Allard2013,bement2024patterning,beta2023actin,inagaki2017actin}.

Mathematical modelling and analysis are often plausible methods to investigate pattern formation mechanisms in natural systems as well as to study hypothesized underlying interactions. However, mathematical models for the emerging ``actin waves'' that couple F-actin and its regulators often tend to be complex, needing to rely on numerical simulations to recapitulate experimental observations~\cite{Allard2013,beta2023actin}. The drawback, however, with studies based solely on simulations is that conclusions obtained may obscure the role of generic emergent behaviours in favour of model details~\cite{Arai2010,PoncedeLeon2021,doubrovinski2011cell}. Moreover, the parameter space of many models is typically very large, so finding biological regimes or mapping out the possible behaviours may be impractical. Even if resemblance to observations can be argued as support for such models, it can be difficult to understand, from simulations alone, how patterns emerge, interact, coexist, and affect each other's stability. Lacking such insights makes it difficult to suggest nontrivial predictions, which are essential to the validation of models. Here, we show that mechanistic understanding can be obtained via a relatively simple ``toy model'' that is amenable to analysis, specifically in the form of spatial dynamics. Examples, in the context of actin waves, where this approach has proven highly informative, include~\cite{bernitt2017fronts,Yochelis2022}. 

\begin{figure}
    \centering
    \includegraphics[width=0.8\linewidth]{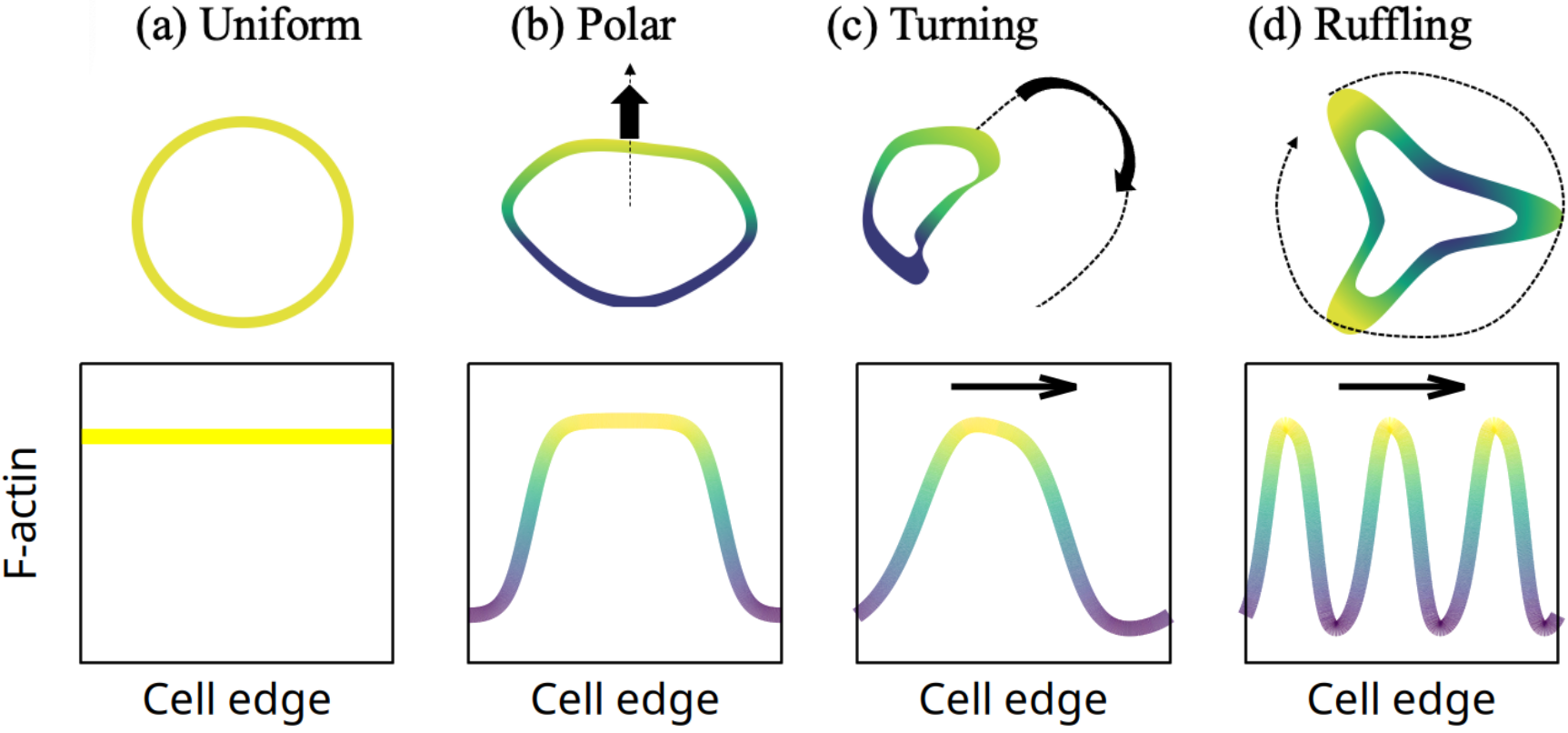}
    \caption{Schematic representation of cells and corresponding F-actin dynamics at the cell edge. Top row: Four cell behaviours showing typical cell shapes and dynamics, together with F-actin distribution along the cell edge; light (dark) colour indicates high (low) F-actin concentration. Bottom row: Corresponding numerical solutions on a one-dimensional (1D) domain with periodic boundary conditions. (a) Stable high uniform F-actin distribution depicting an unpolarized (``resting’’) cell. (b) Polar distribution associated with directed cell migration. (c) Travelling wave (TW) with a single wavelength in the periodic domain. This TW would lead the cell to crawl/turn in a circular arc. (d) Travelling wave with three wavelengths, resulting in a ``ruffling'' behaviour, with three protrusions circulating around its edge.}
    \label{fig:cellbehav}
\end{figure}

Our aim, hence, is to identify a (simple) polynomial-type but relevant {prototypical} model from which generic mechanisms for GTPase-actin dynamics, in the context of cell motility, can be drawn. We pose the following questions:
\begin{description}
    \item [(Q1)] Based on previous works~\cite{brauns2024nonreciprocal,cao2019plasticity,holmes2017mathematical, Jacobs2019,Yochelis2022} that show distinct regimes of patterns in parameter space, is it possible to identify a mechanism for \textit{bistability} of travelling waves and polarization in non-gradient reaction-diffusion models with mass conservation? 
    \item [(Q2)] What are the implications for cell motility modes? Such modes (summarized in~\cref{fig:cellbehav}) include (a) steady and symmetric patterns, (b) polarized patterns, corresponding to directed migration, (c) turning or rotating cells, and (d) ruffling cells, with several protrusions (``lamellipodia'') circulating around their perimeter.
\end{description}
To address the above, we {modify} a three-variable mass-conserving reaction-diffusion model by~Holmes \textit{et al.}~\cite{holmes2012regimes}, in one space dimension (1D) on a ring (periodic cell edge) and use bifurcation analysis to dissect it. 
\\ \\
\noindent The paper is organized as follows: 
\begin{description}
    \item [\Cref{sec:model description},] modification of the non-gradient mass-conserving reaction-diffusion model for actin waves by Holmes \textit{et al.}~\cite{holmes2012regimes}, {by replacing Hill functions with polynomial nonlinearities representing the coupling between the GTPase and F-actin};
    \item [\Cref{sec:LSA},] linear analysis of the homogeneous steady states (HSSs) in two-parameter space: we trace out the locus of distinct bifurcations and identify a codimension-2 point where a steady long wavelength and finite wavenumber Hopf instability emerge simultaneously;
    \item [\Cref{sec:WnonlinA},] weakly nonlinear analysis of the finite wavenumber Hopf bifurcation, explaining the focus on {short travelling wave (TW) trains, i.e., TWs with a small number of spatial wavelengths};
    \item [\Cref{sec:full bifurcation analysis},] numerical bifurcation analysis via the continuation method to investigate the emerging structures identified in~\Cref{sec:LSA}: we track {long wavelength oscillatory (LWO) solutions}, excitable pulses (EPs), travelling fronts (TFs), {stationary pulses (SP)}, and static ``wave-pinning’’ (WP) solutions in large domains, as well as TWs and WP solutions in moderately sized domains;
    \item [\Cref{sec:time simulations},] numerical simulations to investigate large amplitude perturbations and the robustness of TW and WP solutions;
    \item [\Cref{sec:discussion},] conclusion of the findings and outlook.
\end{description}

\begin{figure}[tp]
\centering
\begin{tabular}{lcr}
\begin{subfigure}[htbp]{0.17\textwidth}
\centering
\includegraphics[height=3.9cm]{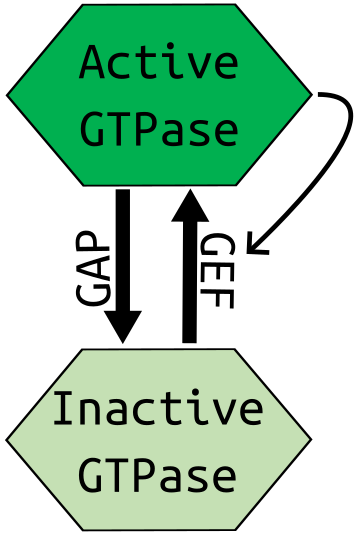}
\caption{\quad\quad\quad\quad\quad\quad\quad}\label{fig:Mori}
\end{subfigure}&
\begin{subfigure}[htbp]{0.37\textwidth}
\centering
\includegraphics[height=3.9cm]{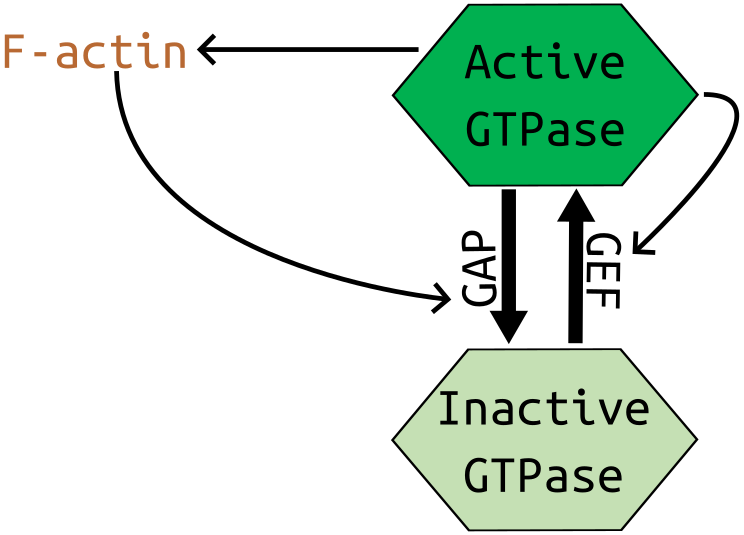}
\caption[]{} \label{fig:Holmes}
\end{subfigure}&
\begin{subfigure}[htbp]{0.37\textwidth}
\includegraphics[height=3.9cm]{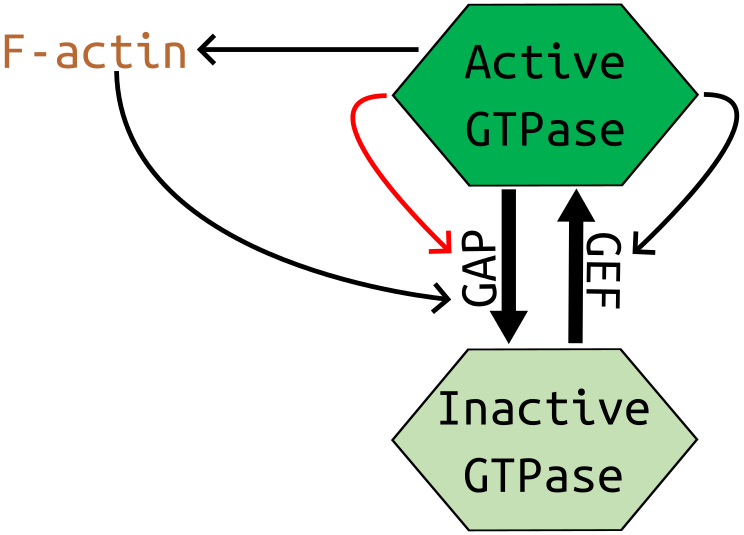}
\caption{} \label{fig:poly}
\end{subfigure}
\end{tabular}
\caption{Schematic diagrams of several Rho-GTPase feedback-loop models. (a) The {Mori \textit{et al.}\ model~\cite{Mori2008} for cell polarization, \cref{eq:MoriModel}, }
comprises a single GTPase with active (dark) and inactive (light) forms that are interconverted by the enzymatic action of a GEF (activation) and a GAP (inactivation). The total amount of GTPase remains constant (mass conservation)~\cite{Mori2008}. The thin arrow represents saturating positive feedback of active GTPase on its own activation. (b) The ``GTPase-actin'' model of Holmes~\cref{eq:HolmesModel} incorporates assembly of F-actin downstream of the GTPase and negative feedback from F-actin (via a GAP)~\cite{holmes2012regimes}. (c) The modified model used throughout this study, see~\Cref{eq:model}, replaces the saturating positive feedback by a combination of polynomial positive and negative (red) feedback, otherwise preserving the structure of~\cref{eq:HolmesModel}.  
}\label{fig:reaction schematic}
\end{figure}

\section{Model equations for the GTPase-actin system}\label{sec:model description}
The starting point of our modelling is the two-component mass-conserved reaction-diffusion system for cell polarization proposed by Mori \textit{et al.}~\cite{Mori2008} (\Cref{fig:Mori}). Essential features of this system are slow-fast rates of diffusion for active ($u$) and inactive ($v$) GTPase, respectively, and nonlinear kinetic terms that admit bistability in some parameter regimes~\cite{goryachev2017many}. On the timescales of interest (seconds, minutes) the GTPase is neither produced nor degraded, so Mori \textit{et al.}~\cite{Mori2008} assumed the total mass of GTPase was conserved. Therefore, the bistability differs from that of generic non-gradient models, as there is a single nullcline and the flow is restricted by mass conservation. In particular, since the system can be written in a gradient form, the steady-states depend on the Maxwell construction and are non-hyperbolic~\cite{bergmann2019system,brauns2020phase,champneys2021bistability,kuwamura2024single,mori2011asymptotic,otsuji2007mass,robinson2025universal}. This system produces a spatially nonuniform steady solution {(denoted a wave-pinning solution)} associated with cell polarity: a high plateau of active GTPase (corresponding to the leading edge of a cell) separated by an interface from a low active GTPase plateau. Note that wave-pinning (WP) solutions are also known as ``mesa'' states in the context of porous media~\cite{elliott1986mesa} and chemical patterning~\cite{kolokolnikov2006mesa}. Holmes \textit{et al.}~\cite{holmes2012regimes} extended the {Mori \textit{et al.}\ model~\cite{Mori2008}} to capture ``actin waves" in reasoning that is similar to the idea behind the classic FitzHugh-Nagumo (FHN) system~\cite{fitzhugh1961impulses,nagumo1962active}, where a bistable system is coupled to slow negative feedback: the $(u,v)$-system acts as the ``bistable" part, and slow negative feedback is provided by F-actin ($F$), whose role is similar to a ``refractory variable'' in the FHN model (\Cref{fig:Holmes}). The experiments of Michaud \textit{et al.}~\cite{Michaud2022} describe a real cell system with interactions as in the Holmes \textit{et al.} model (although the authors chose to rederive their own, rather similar, model system): the GTPase is Rho, it self-activates (via the GEF Ect2), and F-actin recruits a GAP (RGA-3/4) that inactivates Rho. In what follows, we simplify the original WP and actin waves models to produce the model that we will analyze in the remainder of this study. 

The {Mori \textit{et al.}\ model}~\cite{Mori2008} reads in its dimensionless form as:
\begin{subequations}\label{eq:MoriModel}
    \begin{align} 
    \frac{\partial u}{\partial t}&= \left(b+\gamma\frac{u^n}{1+u^n}\right)v -  I u+ D\frac{\partial^2 u}{\partial x^2},\label{eq:Mori u}\\
    \frac{\partial v}{\partial t}&= -\left(b+\gamma\frac{u^n}{1+u^n}\right)v+ I u  +\frac{\partial^2 v}{\partial x^2}.\label{eq:Mori v}
\end{align}
The model conserves mass,
\begin{align} \label{eq:mass con}
M:=\frac{1}{L}\int_0^L\left[u(x,t)+v(x,t)\right]{\textrm d}x=\text{constant},
\end{align}
where $L$ is the domain length.
\end{subequations}
Here $u,v$ are, respectively, the levels of active and inactive Rho-GTPase, $b$ is a basal rate of activation, $I$ is an inactivation rate, $\gamma$ is the rate of autoactivation (positive feedback of active GTPase to its own activation rate via a GEF), and $D<1$ is the slow rate of diffusion of active GTPase relative to the inactive GTPase. (The disparity in diffusion rates stems from the fact that the active GTPase is bound to the membrane of the cell, a more viscous environment than the cytosol where the inactive GTPase resides.) This model arbitrarily assumes positive feedback onto the GEFs, but it was also shown that negative feedback onto the GAPs leads to qualitatively similar results~\cite{jilkine2011comparison}. For $n=2$ and $I=1$ in \cref{eq:MoriModel}, WP solutions are observed over a wide range of parameters. These solutions are interpreted as polarized patterns of active GTPase~\cite{Mori2008,mori2011asymptotic} associated with cell polarity. It is then understood that if $u$ is the level of active Rac or Cdc42 (both Rho-GTPases that promote branched F-actin), then wherever $u$ is high, there would locally be assembly of F-actin that can exert protrusive force on the cell membrane. Alternatively, if $u$ represents the GTPase Rho, then there could be either F-actin assembly (by formins, proteins that nucleate F-actin strands) or local myosin activation that leads to edge contraction~\cite{ridley2015rho}.

Holmes \textit{et al.}~\cite{holmes2012regimes} coupled the {Mori \textit{et al.}\ model~\cite{Mori2008}} to F-actin ($F$) via slow negative feedback, assuming that $F$ increases the rate of inactivation of the GTPase $u$ (see~\Cref{fig:Holmes}). The dynamics of $F$ are taken to be linear and its rate of diffusion, which is very low was neglected:
\begin{subequations}\label{eq:HolmesModel}
    \begin{align}
    \frac{\partial u}{\partial t}&= \left(b+\gamma\frac{u^2}{1+u^2}\right)v -  \left(1+ s\frac{F}{1+F} \right) u+ D{\frac{\partial^2 u}{\partial x^2}},\label{eq:Holmes u}\\
    \frac{\partial v}{\partial t}&= -\left(b+\gamma\frac{u^2}{1+u^2}\right)v+ \left(1+ s\frac{F}{1+F} \right) u  +{\frac{\partial^2 v}{\partial x^2}},\label{eq:Holmes v}\\
    \frac{\partial F}{\partial t}&= \theta (p u - F).\label{eq:Holmes F}
\end{align}
\end{subequations}
We note that conceptually, system~\eqref{eq:HolmesModel} is similar to the actin waves model of~\cite{Yochelis2022}. We later draw comparisons for selected values of the total GTPase concentration $M$. 

{To develop an intuitive understanding of the mechanisms leading to robustness of patterns (here and for future studies), we choose to modify the Holmes model~\cref{eq:HolmesModel} to create a prototypical model, i.e., a model including polynomial terms as in other reaction-diffusion models~\cite{jd2003mathematical,murray2007mathematical}, such as the FHN and Gray-Scott models. Nevertheless, the modification preserves key relevant intracellular features: the positive feedback of $u$ on the assembly of $F$ (taking $p_1\ge 0$) and the slow negative feedback of $F$ on $u$ (see~\Cref{fig:poly}). To avoid unbounded growth, $u$ is assumed to promote its own inactivation. We also add an F-actin basal assembly rate $p_0$ to expand the pattern-forming regime.} The resulting model equations are given by
\begin{subequations}\label{eq:model}
    \begin{align} 
    \frac{\partial u}{\partial t}&= (b+\gamma u^2)v - (1+s F+u^2) u+ D\frac{\partial^2 u}{\partial x^2},\label{eq:model u}\\
    \frac{\partial v}{\partial t}&= -(b+\gamma u^2)v+ (1+s F+u^2) u  +\frac{\partial^2 v}{\partial x^2},\label{eq:model v}\\
    \frac{\partial F}{\partial t}&=\theta (p_0+p_1 u - F) +D_F \frac{\partial^2 F}{\partial x^2}.\label{eq:model F}
\end{align}
\end{subequations}
In what follows, we analyze system~\eqref{eq:model} on a 1D domain with either periodic or Neumann boundary conditions (PBCs or NBCs, respectively), interpreted as a ring corresponding to the cell perimeter (see~\Cref{fig:cellbehav}), or a cell diameter, respectively,
\begin{align} \label{eq:BCs} 
Q(0,t)=Q(L,t)\quad\quad\quad\quad\text{and}\quad\quad\quad\quad \partial_x Q(0,t)=\partial_x Q(L,t)=0, 
\end{align} 
where $Q(x,t):=(u,v,F)^{\rm T}$ and the superscript ${\rm T}$ stands for transpose.

For analysis purposes, we use the strength of F-actin negative feedback, $s$, as the primary bifurcation parameter and the GTPase basal activation rate $b$ as the secondary parameter. See also the complete list of parameter values and their biological meanings in~\Cref{tab:par values}.

\begin{table}[tp]
\footnotesize
\caption{Parameter descriptions and values used throughout this study.} \label{tab:par values}
\begin{center}
    \begin{tabular}{|c|l|c|} \hline
        \bf Parameter & \bf Definition & \bf Value\\ \hline
        $b$ & GTPase basal activation rate & [0,4.5]\\
        $\gamma$ & GTPase autocatalytic activation rate & 3.557\\
        $s$ & Strength of F-actin negative feedback& [0,14]\\
        $\theta$ & F-actin time scale parameter& 0.6\\
        $p_0$ & F-actin basal assembly rate & 0.8\\
        $p_1$ & GTPase-dependent F-actin assembly rate & 3.8\\
        $D$ & Active GTPase rate of diffusion & 0.1\\
        $D_F$ & F-actin rate of diffusion & 0.001 \\ 
        $M$ & Average total GTPase concentration & 2,4.5\\ \hline
    \end{tabular}
\end{center}
\end{table}

\begin{figure}[tp]
    \centering
    \begin{tabular}{cc}
        \begin{subfigure}{0.45\textwidth}
            \includegraphics[width=\textwidth]{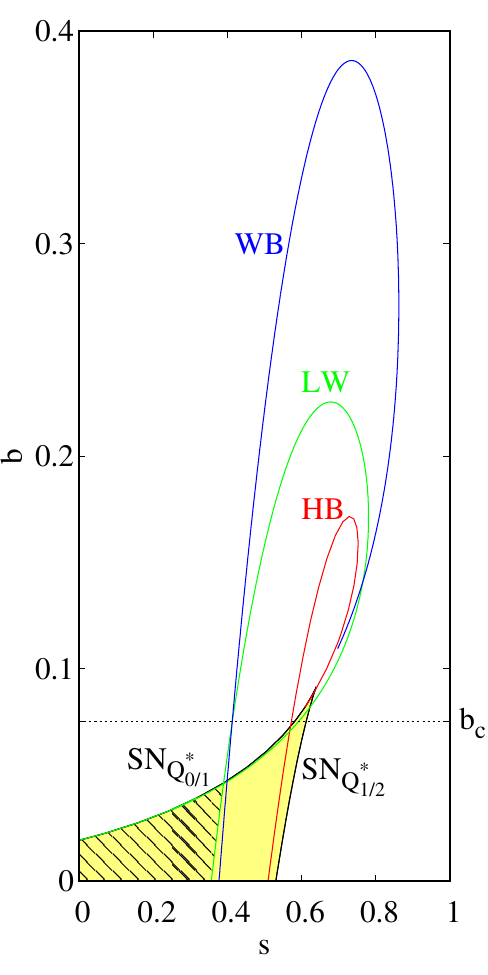}
            \caption{$M=2$} \label{fig:2par bif HSS M2}
        \end{subfigure}&
        \begin{subfigure}{0.45\textwidth}
            \includegraphics[width=\textwidth]{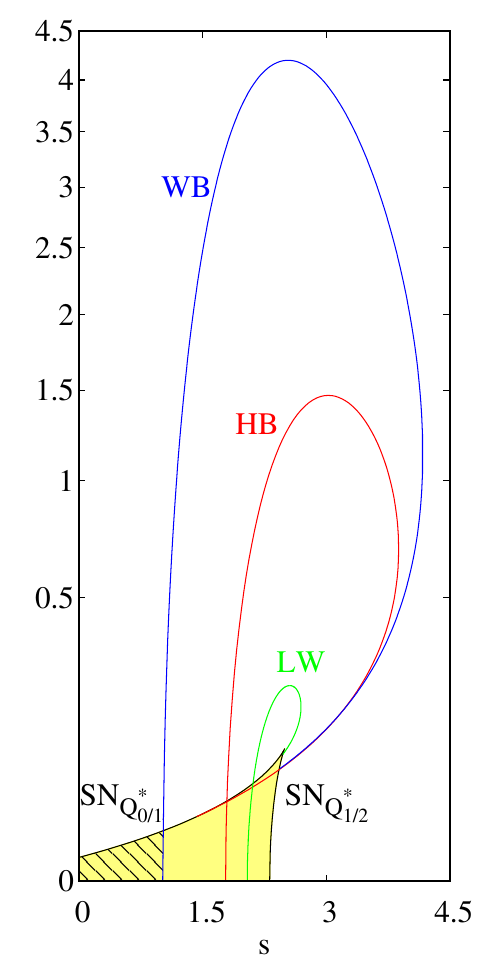}
            \caption{$M=4.5$} \label{fig:2par bif HSS M45}
        \end{subfigure}
    \end{tabular}
    \caption{The linear onsets of bifurcations along homogeneous steady states (HSSs) are shown as curves in the $(s,b)$ parameter plane, where $s$ is the strength of F-actin negative feedback and $b$ is the GTPase basal activation rate. The bifurcation onsets are computed using~\cref{eq:disper eigen}, where LW - long wavelength (green), WB - finite wavenumber Hopf (blue), and HB - homogeneous Hopf (red). (a) Low vs (b) high total GTPase concentration ($M$) differ since in (a) LW occurs before HB while in (b) it is reversed (note that in (b) we use a logarithmic dependence of $b$). 
    {Three HSSs coexist in the entire yellow region. The region is bounded by saddle-nodes SN that end at a cusp bifurcation at (a) $(s,b)\approx(0.64,0.092)$ and (b) $(s,b)\approx(2.5,0.11)$. The black-striped subset of the yellow region corresponds to the bistability region, where two HSSs out of three are stable. In the solely yellow-shaded region, there is at most one stable HSS.} The horizontal dotted line in (a) at $b=b_c\approx0.067$, corresponds to a one-parameter slice shown in~\Cref{fig:2par bif 1par disper}, where a codimension-2 LW/WB instability occurs. {See~\cref{fig:2 par appendix i,fig:2 par appendix ii,fig:2 par appendix iii,fig:2 par appendix v} for more one-parameter slices of~\cref{fig:2par bif HSS}a.} Other parameter values as in~\Cref{tab:par values}.} \label{fig:2par bif HSS}
\end{figure}

\begin{figure}[tp]
\centering
\includegraphics[width=\textwidth]{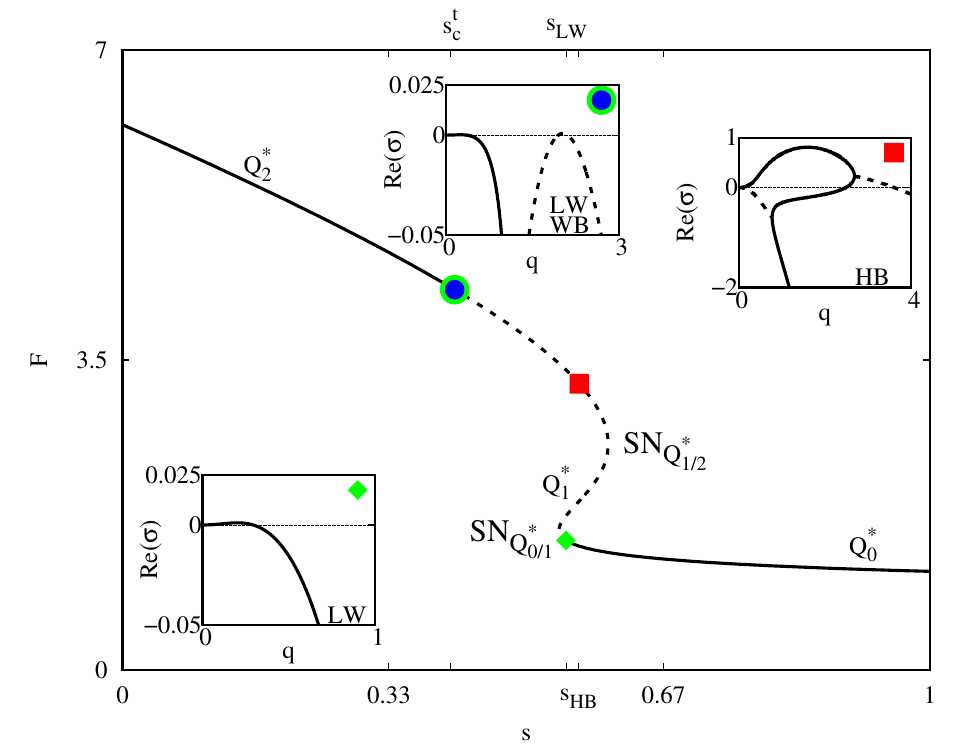}
\caption{One-parameter bifurcation diagram for the F-actin component as a function of the strength of F-actin negative feedback $s$, along the dotted line slice in~\Cref{fig:2par bif HSS M2}, where $Q^\ast_{0,1,2}$ are homogeneous steady states; solid (dashed) lines denote linear stability (instability). The values $(s_c^t,s_{LW},s_{HB})\approx (0.409,0.548,0.565)$ denote the top codimension-2 long wavelength (LW)/finite wavenumber Hopf (WB) instability onset about $Q_2^\ast$, LW instability onset about $Q_0^\ast$, and the homogeneous Hopf (HB) bifurcation, respectively. Insets show dispersion relations at bifurcation onsets indicated by corresponding shapes (circle, square, etc.); the growth rate Re$(\sigma(q))$ of perturbations with wavenumber $q$ is a solid (dashed) line for real (complex conjugate) parts. Parameter values as in~\Cref{tab:par values} with $M=2$ and $b=b_c\approx0.067$.} \label{fig:2par bif 1par disper}
\end{figure}

\section{Linear analysis of homogeneous steady states (HSSs)} \label{sec:LSA}

The HSSs of~\cref{eq:model} (after setting $\partial_t=\partial_x=0$) are solutions to
\begin{subequations}\label{eq:HSS equations}
    \begin{align}
        0&=(b+\gamma u^2)v - (1+s F+u^2) u,\\
        0&=\theta (p_0+p_1 u - F),\\
        0&=u+v-M.
    \end{align}
\end{subequations} 
The third equation comes from mass conservation.
Elimination of $v$ and $F$, $v=M-u$ and $F=p_0+p_1u$, leads to
\begin{equation} \label{eq:HSS u}
    \nu(u)\equiv(b+\gamma u^2)(M-u) - (1+s(p_0+p_1u)+u^2)u=0.
\end{equation}
At least one biologically relevant HSS $Q^\ast=(u^\ast,v^\ast,F^\ast)^{\rm T}$ (i.e.\ $u^\ast,v^\ast,F^\ast>0$) exists for $b,\gamma,s,p_0,p_1,M>0$ by the \textit{Intermediate Value Theorem} after noting that
\begin{align*}
    \nu(0)&=bM>0,\\
    \nu(M)&=-(1+s(p_0+p_1M)+M^2)M<0.
\end{align*}
Note that $b>0$ excludes the existence of trivial HSSs. In~\Cref{fig:2par bif HSS}, we show the regions where these solutions coexist, i.e., $Q^\ast_{0,1,2}$, where the yellow shaded region corresponds to the existence of multiple steady states and is limited by two folds (black lines) that emanate from a cusp bifurcation. The cross-hatched yellow region denotes bistability of HSSs.

Next, we perform a linear stability analysis of infinitesimal perturbations in the infinite domain by expanding
\[
Q=Q^\ast+\epsilon Q_1+\mathcal{O}(\epsilon^2),
\]
where $|\epsilon|\ll1$ and $Q_1$ is of $\mathcal{O}(1)$. The function $Q_1$ is given by
\[
Q_1= \left(\begin{array}{c}
    u_1\\
    v_1\\
    F_1
    \end{array}\right) e^{\sigma t+iqx}+c.c.,
\]
where $\sigma$ is the growth rate of the wavenumber $q$, and $c.c.$ stands for complex conjugate.
Consequently, linearization leads to
\begin{align} \label{eq:disper eigen}
    \sigma \left(\begin{array}{c}
    u_1\\
    v_1\\
    F_1
    \end{array}\right)=\mathcal{L}(Q^\ast)\left(\begin{array}{c}
    u_1\\
    v_1\\
    F_1
    \end{array}\right),
\end{align}
where 
\begin{equation}\label{eq:linoperator}
    \mathcal{L}{(Q^\ast)}=\left(
    \begin{array}{ccc}
    2 \gamma u^{*} v^{*}-1 - F^{*} s - 3 u^{*2}  - D q^2 & b + \gamma u^{*2} & -s u^\ast \\
    1 + F^{*} s + 3 u^{*2}- 2 \gamma u^{*} v^{*} & -b - \gamma u^{*2}-q^2 & s u^\ast \\
 \theta p_1 & 0 & -\theta-D_Fq^2 \\
\end{array}
\right)
\end{equation}
{is the Jacobian of~\cref{eq:model} evaluated at the HSS $Q^\ast$.} By solving $\det[{\mathcal{L}{(Q^\ast)}-\sigma \mathbb{I}}]=0$ for $\sigma(q)$, where $\mathbb{I}$ is the identity matrix, we obtain three dispersion relations from which instability onsets are classified~\cite{cross1993pattern}. Due to mass conservation, for $q=0$, we expect a persistent zero eigenvalue since $\mathcal{L}(Q^\ast)$ is singular. Linear stability of $Q^\ast$ is, therefore, determined if all growth rates $\sigma(q)$ have negative real parts $\forall q\geq0$, whereas the instability onsets occur once $\text{Re}(\sigma(q_c))=0$ for some {$q_c\geq0$}. If additionally, $\omega_c \equiv \text{Im}(\sigma(q_c))\neq0$, then the instability is of oscillatory type so that two dispersion curves are complex conjugates.

In our setting, we are mainly interested in two instabilities: the steady long wavelength (LW) and the finite wavenumber Hopf (WB). The LW occurs as in gradient systems, i.e., this instability occurs when the {curvature} of $\max_{\sigma}\text{Re}(\sigma(q))$ changes sign at $q=0$~\cite{cross1993pattern}, since at least one $\sigma(0)=0$ by mass conservation~\cite{bergmann2018active}. The WB is an oscillatory type, simultaneously giving rise to both travelling waves (TW) and standing waves (SWs)~\cite{knobloch1986oscillatory}. In addition, we find it informative to monitor bifurcations occurring when the HSS is linearly unstable, i.e., not only the LW and the WB but also the homogenous Hopf (HB). Consequently, we show in \Cref{fig:2par bif HSS}, selected two-parameter diagrams while varying $(s,b)$, for $M=2$ and $M=4.5$, in which we identify saddle-nodes (SNs) of HSS (black) and the locus of bifurcation onsets: HB (red), WB (blue), and LW (green). For an illustration of the bifurcation onsets and the following analysis, we focus on a selected slice at $b=b_c\approx0.067$ (dotted line in~\Cref{fig:2par bif HSS M2}) exhibiting a codimension-2 point of LW/WB instabilities for $M=2$, as shown in~\Cref{fig:2par bif 1par disper}. Here we have two LW bifurcations, where the second is near the $SN_{Q_{0/1}^\ast}$. {In~\Cref{app:slices}, we show additional slices where more codimension-2 points exist (see~\cref{fig:2 par appendix i,fig:2 par appendix ii,fig:2 par appendix iii,fig:2 par appendix v}).} We focus on $M=2$ since for $M=4.5$, the LW loop is the most inner one (see~\Cref{fig:2par bif HSS M45}), excluding the occurrence of the LW/WB instability. Nevertheless, we briefly demonstrate the similarities and differences to the case of $M=4.5$ (in~\Cref{sec:full bifurcation analysis}) and compare with the results obtained by Yochelis \textit{et al.}~\cite{Yochelis2022}, see also~\Cref{sec:Yochelis model}. We note that for $b\approx0.14$, in a region above the cusp bifurcation of HSSs (shaded yellow), there is another codimension-2 LW/WB bifurcation. The latter is outside the scope of this work and will be addressed elsewhere.

\section{{Bifurcation criticality of single-wavelength oscillatory solutions}} \label{sec:WnonlinA}
{We proceed here with the analysis of the bifurcation criticality and relative linear stability of oscillatory solutions emerging from the LW/WB instability, $s=s_{WB}=s_c^t\approx0.409$ and $b=b_c\approx0.067$, up to larger values of $b$ and about $s=0.6$, see~\Cref{fig:2par bif HSS M2}. At the LW/WB onset, the steady solutions bifurcate subcritically (see~\Cref{fig:EP codim2 bif}), namely toward the linearly stable regime, while the travelling ($\text{TW}_{\lambda}$) and standing ($\text{SW}_{\lambda}$) waves with critical wavelength $\lambda=2\pi/q_c\approx3.093$ are supercritical with $\omega_c\approx 0.388$~\cite{hughes2025dissipative}. To check if the criticalities of the single-wavelength TWs and SWs remain supercritical {for the varied parameter values}, we perform weakly nonlinear stability analysis, a.k.a.\ employing multiple timescales or the amplitude equation method. We ignore weak spatial modulations, including the impact of the large-scale mode~\cite{matthews2000conservation,winterbottom2005oscillatory}, due to the relationship to cells with perimeters comparable to a small number of spatial wavelengths.}

Let $\epsilon$ be a small parameter such that $\epsilon^2 \propto |s-s_{WB}|\ll1$ is a measure of the system's distance from the onset of instability. {Then $s = s_{WB} + \epsilon^2 \widetilde s$, where $\widetilde s$ is of $\mathcal{O}\left(1\right)$ and respectively, $Q^\ast = Q_c^\ast(s_{WB}) + \epsilon^2 \widetilde Q^\ast$}. We set the power of $\epsilon$ in the slow temporal scale by observing the dependence of the growth rate $\sigma$ on $\epsilon$. Finding a quadratic dependence, $\sigma\propto\epsilon^2$, we arrive at the temporal scaling $\tau=\epsilon^2 t$. {The amplitudes for the left/right propagating waves (slowly varied in time), are obtained by expanding $Q$ in powers of $\epsilon$,
\begin{subequations}\label{eq:perturb}
	\begin{align}
	u &= u^\ast_c+\epsilon^2 \widetilde u^\ast + \epsilon u_1(x,t,\tau) + \epsilon^2 u_2 + \mathcal{O}\bra{\epsilon^3},\\
	v &= v^\ast_c+\epsilon^2 \widetilde v^\ast + \epsilon v_1(x,t,\tau) + \epsilon^2 v_2 + \mathcal{O}\bra{\epsilon^3},\\
	F &= F^\ast_c+\epsilon^2 \widetilde F^\ast + \epsilon F_1(x,t,\tau) + \epsilon^2 F_2 + \mathcal{O}\bra{\epsilon^3},
	\end{align}
\end{subequations}
and using the ansatz:
\begin{subequations}\label{eq:firstorder}
	\begin{align}
	   u_1 &= B_{Lu}(\tau)e^{i(\omega_ct + q_cx)} + B_{Ru}(\tau)e^{i(\omega_ct - q_cx)} + c.c.,\\
	   v_1 &= B_{Lv}(\tau)e^{i(\omega_ct + q_cx)} + B_{Rv}(\tau)e^{i(\omega_ct - q_cx)} + c.c.,\\
	   F_1 &= B_{LF}(\tau)e^{i(\omega_ct + q_cx)} + B_{RF}(\tau)e^{i(\omega_ct - q_cx)} + c.c..
	\end{align}
\end{subequations}
In~\eqref{eq:firstorder}, $\omega_c$ and $q_c$ denote the critical frequency and wavenumber at the instability onset, and subscripts $L,R$ stand for left and right, respectively. Following a standard procedure (detailed in~\Cref{sec:amp_eq_app}), leads to the equations:  
\begin{subequations}\label{eq:amps}
    \begin{align}
        \dot{B}_{LF} &= \alpha (s - s_{WB}) B_{LF} - \bra{\gamma |B_{LF}|^2 + \eta |B_{RF}|^2 } B_{LF},\\
        \dot{B}_{RF} &= \alpha (s - s_{WB}) B_{RF} - \bra{\gamma |B_{RF}|^2 + \eta |B_{LF}|^2 } B_{RF}.
    \end{align}
\end{subequations}}
\begin{figure}[tp!]
    \centering
    \includegraphics[width=\textwidth]{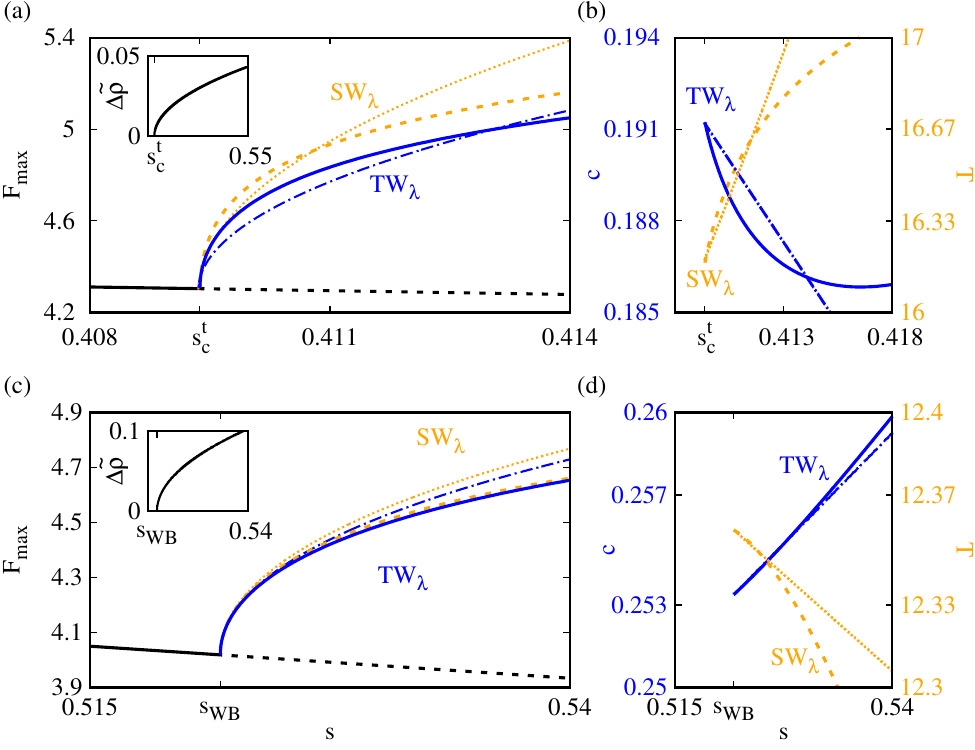}
    \caption{(a) Bifurcating branches of the primary travelling and standing waves (with wavelength $\lambda\approx3.09$) obtained at the codimension-2 onset at $s=s_{WB}=s_c^t\approx 0.409$ and $b=b_c\approx 0.067$ (see the slice in~\Cref{fig:2par bif HSS M2}), hereafter $\text{TW}_{\lambda}$ and $\text{SW}_{\lambda}$, respectively. The solutions are projected onto the maximum F-actin concentration, $F_{\max}$. We compare results from the amplitude equations~\cref{eq:F-field} and~\cref{eq:rel_KnoblochAmps} to numerical continuation, over spatial wavelength $\lambda$ and, in the case of SW, also over the temporal period $T$ (also obtained at the onset): Linearly stable TWs (from continuation: solid blue; from amplitude equations: dashed-dotted blue) and linearly unstable SWs (from continuation: dashed orange; from amplitude equations: dotted orange). The inset shows the difference between the total amplitudes of the TWs and SWs according to~\cref{eq:rel_KnoblochAmps} ($\Delta\tilde\rho={\widetilde \rho}_{TW}-{\widetilde \rho}_{SW}$), indicating that TWs are larger than SWs. (b) Same as (a) but with TWs being projected onto the phase wave speed $c$ ({i.e., the initial propagation speed,} left vertical axis) and SWs onto the time-period $T$ (right vertical axis), see~\cref{eq:c_T_amp_eq} for the corresponding expressions from the amplitude equations. Coefficients for the amplitude equations are obtained numerically using \textit{Mathematica}: $\alpha = 2.13528 - 1.66874i$, $\gamma = 0.06019 + 0.01434i$, $\eta = 0.06500 + 0.16646i$, $\widehat a \approx -0.00482-0.15211 i$, and $\widehat b \approx -0.06019 -0.01434i$. Panels (c,d) provide the same comparison but at $b=0.25$, where the finite wavenumber Hopf onset occurring at $s=s_{WB}\approx 0.522$ with a critical wavelength $\lambda \approx3.13$, and respective coefficients: $\alpha = 1.26708 - 0.46261i$, $\gamma = 0.14570 - 0.12758i$, $\eta = 0.38413 - 0.11440i$, $\widehat a \approx -0.23843 -0.01318i$, and $\widehat b \approx -0.14570 + 0.12758i$. Other parameters as in~\Cref{tab:par values} with $M=2$.} \label{fig:codim2 wave onset}
\end{figure}

Analysis of these equations is convenient using polar forms $B=\rho\exp{i\Phi}$ with the subscripts $re/im$ denoting the real/imaginary parts of coefficients, and for convenience, rewrite in a canonic form of total amplitudes~\cite{knobloch1986oscillatory}
\begin {subequations}\label{eq:KnoblochAmps}
    \begin{align}
        \dot \rho_L &= \rho_L \Bra{\alpha_{re} \bra{s-s_{WB}} + \widehat a_{re} \rho_R^2 + \widehat b_{re} \bra{ \rho_L^2 + \rho_R^2 } },\\
        \dot \rho_R &= \rho_R \Bra{\alpha_{re} \bra{s-s_{WB}} + \widehat a_{re} \rho_L^2 + \widehat b_{re} \bra{ \rho_L^2 + \rho_R^2 } },\\
        \dot \Phi_L &= \alpha_{im} \bra{s-s_{WB}} + \widehat a_{im} \rho_R^2 + \widehat b_{im} \bra{ \rho_L^2 + \rho_R^2 } ,\\
        \dot \Phi_R &= -\alpha_{im} \bra{s-s_{WB}} - \widehat a_{im} \rho_L^2 - \widehat b_{im} \bra{ \rho_L^2 + \rho_R^2 },
    \end{align}
\end{subequations}
{where $\widehat a_{re}\equiv \gamma_{re} - \eta_{re}$, $\widehat a_{im}\equiv \gamma_{im} - \eta_{im}$, and $\widehat b\equiv -\gamma$.}
Steady solutions to~\cref{eq:KnoblochAmps} are
\begin{subequations}\label{eq:uniformamps}
\begin{align}
        \rho &= \sqrt{\frac{\alpha_{re}}{\widehat b_{re}}\bra{s_{WB}-s}}, \quad {\rm for\,\, TWs\,\, with\,\,} (\rho_L,0)=(\rho,0) {\rm\,\,or\,\,} (0,\rho_R)=(0,\rho),\\
        \rho &= \sqrt{\frac{\alpha_{re}}{\widehat a_{re}+2\widehat b_{re}}\bra{s_{WB}-s}}, \quad {\rm for\,\, SWs\,\, with\,\,} (\rho_L,\rho_R)=(\rho,\rho),
\end{align}
\end{subequations}
and thus, the maximal values of $F$ for TWs and SWs, respectively, are
\begin{equation}\label{eq:F-field}
        F_{\rm max} = F^\ast + \begin{cases}
            2 \rho,\quad \rm for\,\, TWs;\\
            4 \rho,\quad \rm for\,\, SWs.
        \end{cases}
\end{equation}
A comparison of the TWs and SWs amplitudes vs.\ numerical continuation is shown in~\Cref{fig:codim2 wave onset}a {(see~\Cref{sec:cont AUTO}} for brief details of the numerical continuation methods).
For consistency, we also replot in the inset of~\Cref{fig:codim2 wave onset}a, the difference in total amplitudes
\begin{equation}\label{eq:rel_KnoblochAmps}
 \widetilde \rho=\sqrt{\rho_L^2 + \rho_R^2}=\begin{cases}
            \sqrt{\dfrac{\alpha_{re}}{\widehat b_{re}}\bra{s_{WB}-s}},\quad \rm for\,\, TWs;\\ \\
            \sqrt{\dfrac{2\alpha_{re}}{\widehat a_{re}+2\widehat b_{re}}\bra{s_{WB}-s}},\quad \rm for\,\, SWs,
        \end{cases}
\end{equation}
of the TWs and SWs (as opposed to plotting in terms of $F_{\rm max}$) to show that the TW branch in terms of $\Delta{\widetilde \rho}={\widetilde \rho}_{TW}-{\widetilde \rho}_{SW}>0$, is indeed larger than the branch of SWs, as expected by a standard theory~\cite{knobloch1986oscillatory}. This also agrees with the linear stability of TWs. For completeness, we also compute the frequency correction
\begin{align}
    \Omega_{TW} &= \bra{\alpha_{im} - \alpha_{re}\frac{\widehat b_{im}}{\widehat b_{re}}}\bra{s-s_{WB}},\\
    \Omega_{SW} &= \bra{\alpha_{im} - \alpha_{re}\frac{\widehat a_{im} + 2 \widehat b_{im}}{\widehat a_{re} + 2 \widehat b_{re}}}\bra{s-s_{WB}},
\end{align}
and the modified speed and temporal period
\begin{equation}\label{eq:c_T_amp_eq}
    c_{TW} = \frac{\omega_c + \Omega_{TW}}{q_c},\quad T_{SW} = \frac{2\pi}{\omega_c + \Omega_{SW}}.
\end{equation}
\Cref{fig:codim2 wave onset}b compares the speed of TWs and temporal period for SWs from \eqref{eq:c_T_amp_eq} with those obtained via numerical continuation.

{Conducting the above computations for the left side of the WB loop, i.e., $b>b_c$, we find that the criticalities of TW and SW persist. As can be seen in~\Cref{fig:codim2 wave onset}, the results derived from amplitude equations agree well far from the codimension-2 LW/WB instability (bottom panels), whereas near the codimension-2 LW/WB instability only qualitatively properties agree (top panels). This is because for $s>s^t_c$, the instability is for both the LW and the oscillatory modes so that the coupling between them affects the supercritical bifurcating oscillatory states. Interestingly, the speed of TWs near the codimension-2 LW/WB instability first decreases and then increases as opposed to only increasing in the absence of the LW onset, as shown in~\Cref{fig:codim2 wave onset}. {Naturally, in the absence of spatial modulations, the derived amplitude equations cannot address this inconsistency.} We also note that below the codimension-2 point, i.e., $b<b_c$, both TW and SW are initially unstable since there, WB lies inside the LW loop (i.e.\ linearly unstable regime of $Q_2^\ast$). Nevertheless, TWs of critical period or otherwise may still gain stability through other mechanisms far from the onset. Some of these will be demonstrated in the following section.}

\section{Solutions and bifurcations in the nonlinear regime} \label{sec:full bifurcation analysis} 

Linear stability analysis reflects the spatial separation of scales associated with the primary long wavelength (LW) and finite wavenumber Hopf (WB) instabilities. Therefore, in what follows, we find it useful to separate the analysis and investigate solutions in large domains ($L\gg \lambda$) and in moderate domains [$L\sim {\mathcal O}(m\lambda$)] along the slice depicted in~\Cref{fig:2par bif 1par disper}, where $\lambda$ is the critical wavelength of the WB in the codimension-2 LW/WB bifurcation at $(s,b)=(s_c^t,b_c)$ and $m$ is some small positive integer. The analysis is performed using {the} numerical continuation package AUTO~\cite{Doedel_2009}. For presentation purposes, we plot the bifurcation diagrams with a combination of the standard Sobolev $H^1$ norm
\begin{align} \label{eq:Sobolev norm}
\norm{Q}:=\sqrt{\frac{1}{L}\int_0^L\left[Q^{\rm T}Q+Q'^{\rm T}Q'\right]{\rm d}\xi},
\end{align}
and/or with the maximum value of $F$ over the domain, and/or the propagation speed $c$, where primes denote differentiation with respect to the argument. All the numerical computations involve steady-state solutions since the TWs are computed in a comoving frame, $\xi=x-ct$. Note that since the solutions of interest obey left-right translational symmetry, we only plot branches for speed $c>0$. For details on the implementation in AUTO, on linear stability computation, and on the direct PDE numerical integration methods, we refer the reader to~\Cref{app:computational implementation}.
\begin{figure}[tp!]
    \centering
    \includegraphics[width=\textwidth]{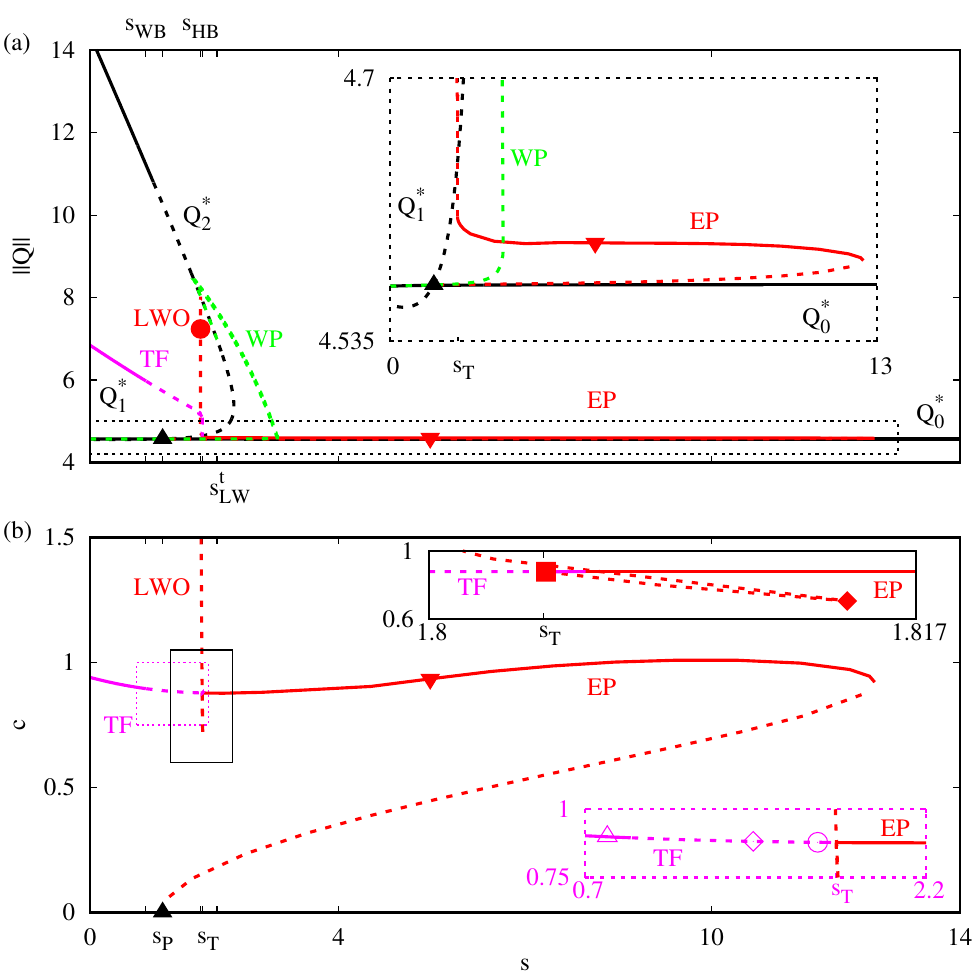}\\
    \caption{Bifurcation diagrams computed via numerical continuation as a function of $s$, showing excitable pulses (EPs), travelling fronts (TFs), wave-pinning (WP) solutions, homogeneous steady states $Q^\ast_{0,1,2}${, and the long wavelength oscillatory (LWO) solutions that emerge from the homogeneous Hopf bifurcation (HB)} for $M=4.5$ and $L=1000\gg \lambda\approx{1.78}$ (where $\lambda$ is the wavelength leading to the instability of $Q_2^\ast$ {in this regime}). The solution branches are projected with the Sobolev norm~\cref{eq:Sobolev norm} in (a) and the propagation speed $c$ in (b); solid (dashed) lines denote linearly stable (unstable) solutions. The inset in (a) zooms into the stable region of EPs while the insets in (b) zoom into the T-point regime. The critical values $(s_{WB},s_P,s_{HB},s_T,s_{LW}^t)=({1.023},1.167,1.781,1.804,2.044)$ represent the finite wavenumber Hopf (WB) instability of $Q_2^\ast$, the parity-breaking bifurcation (black triangle) of {stationary pulses} to EPs, the homogeneous Hopf (HB) bifurcation onset, the T-point (red square), and the top ($Q_2^\ast$) long wavelength (LW) bifurcation, respectively. {After the T-point, $s=s_T$, EPs bi-asymptotic to $Q_2^\ast$ become bi-asymptotic to $Q^\ast_1$, and then LWO solutions after an additional homoclinic bifurcation (for details, see~\cite{Yochelis2022}) with a speed of $c\approx 410$ at $s=s_{HB}$ but is cropped in (b)}. The symbols (circle, diamond, etc.)\ represent selected locations of solution profiles given in~\cref{fig:EP M45 solutions}. Other parameter values as in~\Cref{tab:par values} with $b=0.00067$.}\label{fig:EP M45 bif}
\end{figure}

\begin{figure}[tp]
    \centering
    \includegraphics[width=\textwidth]{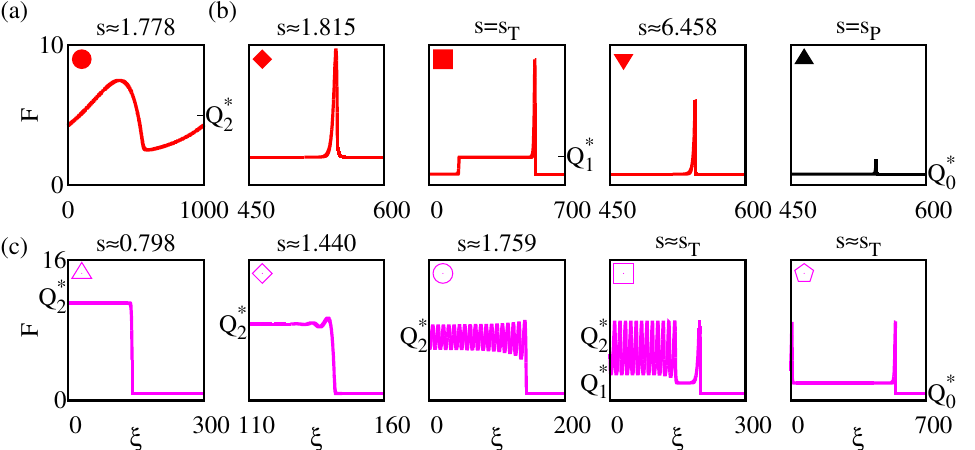}\\
    \includegraphics[width=\textwidth]{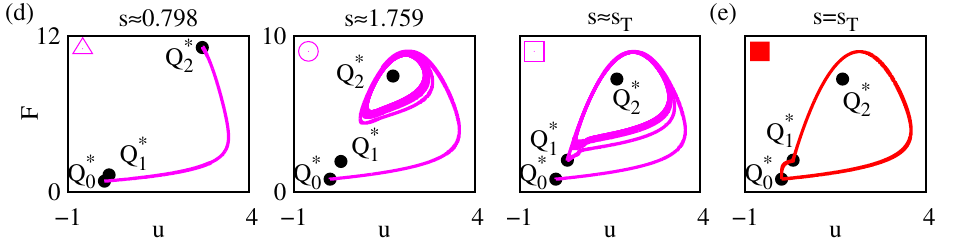} 
    \caption{Solution profiles at selected locations in~\cref{fig:EP M45 bif,fig:EP M45 T-point} along the (a) {LWO}, (b) EP, and (c) TF branches, where for visibility, the whole domain length of $L=1000$ is not always shown. Here $(s_T,s_P)\approx(1.804,1.167)$ are defined as in~\cref{fig:EP M45 bif}. (c,d) Selected phase space projections {(in the comoving frame, $c>0$)} of the (d) TFs and (e) EPs. Note that for all solutions $u,v,F\geq 0$.
    }
    \label{fig:EP M45 solutions}
\end{figure}

\subsection{Coherent solutions in large domains, $L\gg \mathcal{O}(\lambda)$}\label{sec:other patterns full}

The linear stability analysis of $Q_2^\ast$ with $M=2$ {shows} that the instability onset at $s=s_c^t$ is the codimension-2 LW/WB bifurcation (see~\cref{fig:2par bif 1par disper}) while the homogeneous Hopf bifurcation (HB) occurs for $s>s_c^t$. This order of bifurcations differs from the one discussed in~\cite{Yochelis2022}. There, the LW occurs after the HB, which resembles the case with $M=4.5$ (see~\Cref{fig:2par bif HSS M45}). {As in~\cite{Yochelis2022}, the emerging long wavelength oscillatory (LWO) solutions from the HB form into excitable pulses after a homoclinic bifurcation and stabilize at a (global) T-point bifurcation (a codimension-2 bifurcation at which two distinct fronts are simultaneously present (a heteroclinic cycle in a spatial dynamics description), where transitions between homoclinic and heteroclinic orbits occur~\cite{champneys2007shil,glendinning1986Tpoint,knobloch2023front,moreno2022bifurcation,or2001pulse,raja2023,romeo2003stability,sneyd2000traveling,zimmermann1997pulse}).} Therefore, we focus first on coherent solutions for $M=4.5$ using a domain length of $L=1000\gg\lambda\approx {3.09}$ to approximate an infinite domain. The coherent solutions include excitable pulses (EPs) and travelling fronts (TFs), both propagating as spatially localized states corresponding to homoclinic and heteroclinic connections in phase space, respectively, and stationary pulse (SP) and wave-pinning (WP) solutions. Then, we show the fundamental difference between the $M=4.5$ and $M=2$ cases, concerning the instability and stability of WP solutions, respectively.

We demonstrate first the results for $M=4.5$, $L=1000$, and $b=0.00067$, in~\Cref{fig:EP M45 bif}, noting that we confirmed persistence at other values of $b$. The instability of the HSS $Q_2^\ast$ is a finite wavenumber Hopf bifurcation (WB at $s=s_{WB}$). As $s$ increases, subsequent bifurcations are observed: homogeneous Hopf (HB at $s=s_{HB}$), followed by a long wavelength (LW at $s=s_{LW}^t$). As in~\cite{Yochelis2022} and in~\Cref{sec:Yochelis model}, near the HB, {unstable long wavelength oscillatory (LWO) solutions} {emerge with a wavelength matching the domain size (circle). They become} EPs bi-asymptotic in space to $Q_1^\ast$ (i.e., as $\xi\to \pm \infty$) through a homoclinic bifurcation (diamond), then become stable EPs to $Q_0^\ast$ after the global T-point bifurcation (at $s=s_T$, square), and finally, after an additional fold, terminate as unstable EPs at the parity-breaking bifurcation of {SP} states ($s=s_P$, triangle), {for more details, see~\cref{fig:EP codim2 bif}}. The solution profiles {of the LWO and EP solutions} are shown in~\Cref{fig:EP M45 solutions}a{,b, respectively}. {Note that the EPs shown here are saddle connections (in space).} The WP solutions bifurcate subcritically about $Q_2^\ast$ and $Q_0^\ast$ (at $s=s^t_{LW}\approx 2.044$ and $s=s^b_{LW}\approx -0.661$, respectively), and remain linearly unstable everywhere. 

\begin{figure}[h!]
    \centering
    \includegraphics[width=\textwidth]{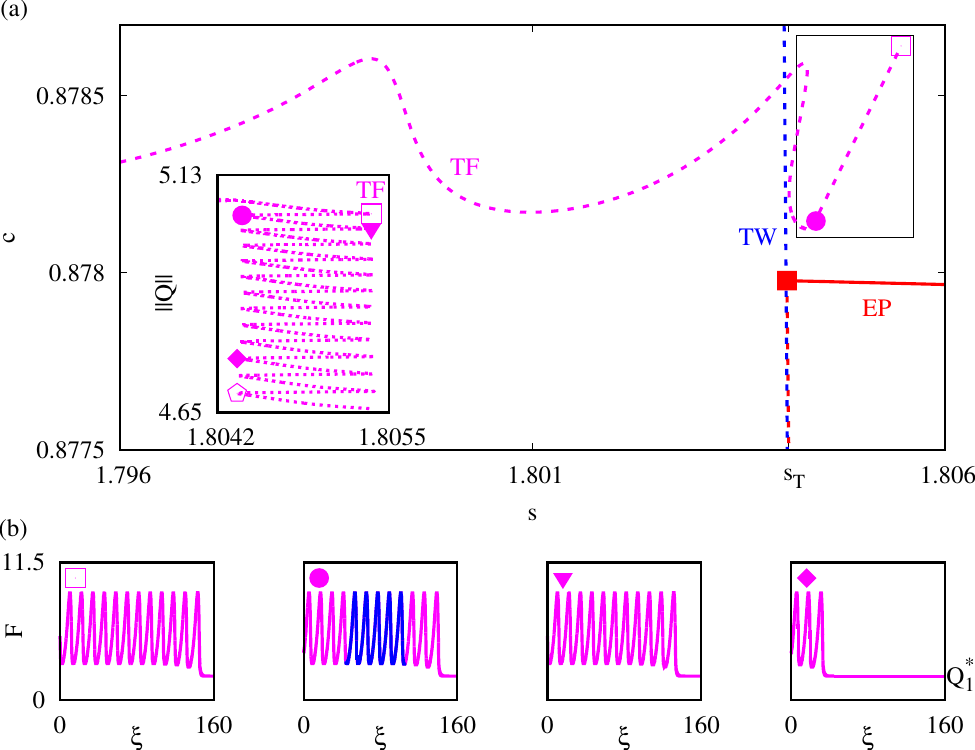}
    \caption{{(a) Bifurcation diagrams computed via numerical continuation as a function of $s$, showing excitable pulses (EPs), travelling fronts (TFs), and travelling waves (TWs), for $M=4.5$ and $L=1000$ near the T-point (square) shown in~\cref{fig:EP M45 bif}. The wavelength of the TWs equals that of the oscillations seen in the periodic-$Q_1^\ast$-$Q_0^\ast$ connections. The solution branches are projected onto the propagation speed $c$; solid (dashed) lines denote linearly stable (unstable) solutions. The critical value $s_T\approx1.804$ represents the location of the T-point involving EP solutions. The inset in (a) shows the region where periodic-$Q_1^\ast$-$Q_0^\ast$ are observed but is instead projected with the Sobolev norm~\cref{eq:Sobolev norm}. Here, we see a snaking behaviour. (b) Selected solution profiles corresponding to the points marked in (a) zoomed in to the spatial oscillations at the left boundary. Each solution shown is a periodic-$Q_1^\ast$-$Q_0^\ast$ cycle (see the hollow square and pentagon in~\cref{fig:EP M45 solutions}c for the full profiles). Here we see that at each fold the spatial oscillations lose half a period. On the circle profile, we overlay the TW solution with five wavelengths showing that the spatial oscillations are travelling waves when isolated on a periodic domain. Note that the TW and TF branches should intersect at the circle but do not because of numerical inaccuracies. Other parameter values as in~\Cref{tab:par values} with $b=0.00067$.}}\label{fig:EP M45 T-point}
\end{figure}

For completeness, we also indicate that in the regime $s^b_{LW}<s<s_{WB}$, there is bistability between the HSSs $Q_0^\ast$ and $Q_2^\ast$ (see~\Cref{fig:EP M45 bif}), which gives rise to TFs. Below the WB instability, $s_{WB}$, fronts connecting $Q_0^\ast$ and $Q_2^\ast$ exhibit monotonic bi-asymptotic spatial decay and are linearly stable, whereas for $s>s_{WB}$, the fronts are unstable while exhibiting spatially decaying oscillations about $Q_2^\ast$. Above $s_{WB}$, these oscillations extend, forming a heteroclinic connection between a spatially periodic orbit and $Q_0^\ast$. {Note that TFs between periodic orbits and homogeneous steady states have been observed in reaction-diffusion systems via numerical integration on a finite domain~\cite{sherratt1995chaos,sherratt2009locating}. Additional increase in $s$ leads to a snaking behavior (see~\cref{fig:EP M45 T-point}), in the vicinity of $s=s_T$, where the solutions resemble double-heteroclinic states (in the comoving frame): periodic-$Q_1^\ast$-$Q_0^\ast$ (see~\cref{fig:EP M45 solutions}c). The periodic orbits in the snaking region are TWs and in~\cref{fig:EP M45 T-point}, we also superimpose a TW branch with a wavelength matching that of the periodic orbits.
Along the snaking branches, the propagation speeds of the solutions oscillate as the periodic orbit is pushed out of the (finite) domain. The solutions lose one period of the left oscillations for every complete cycle from circle to hollow square back to circle in~\cref{fig:EP M45 T-point}. The inset in~\cref{fig:EP M45 T-point} shows a snaking structure observed using the Sobolev norm projection~\cref{eq:Sobolev norm}. At the initial fold (hollow square) there are twelve full periods of the oscillation (with this domain length $L=1000$). At the next fold (circle), half a period is lost, and at the next (inverted triangle), there are only eleven wavelengths. This process continues until only half a period remains (hollow pentagon). In the infinite domain, the reverse process of adding more periods at each fold (i.e., travelling up the branch) would occur indefinitely. Since TFs are not spatially reversible solutions, only one period (not two) is added after passing two saddle-nodes, bearing a similarity to~\cite{ponedel2016forced,li2025traveling}. We demonstrate two profiles in~\Cref{fig:EP M45 solutions}c (see hollow square and pentagon), and in~\Cref{fig:EP M45 solutions}d, we show the respective phase portrait projections.}

\begin{figure}[tp]
    \centering
    \includegraphics[width=\textwidth]{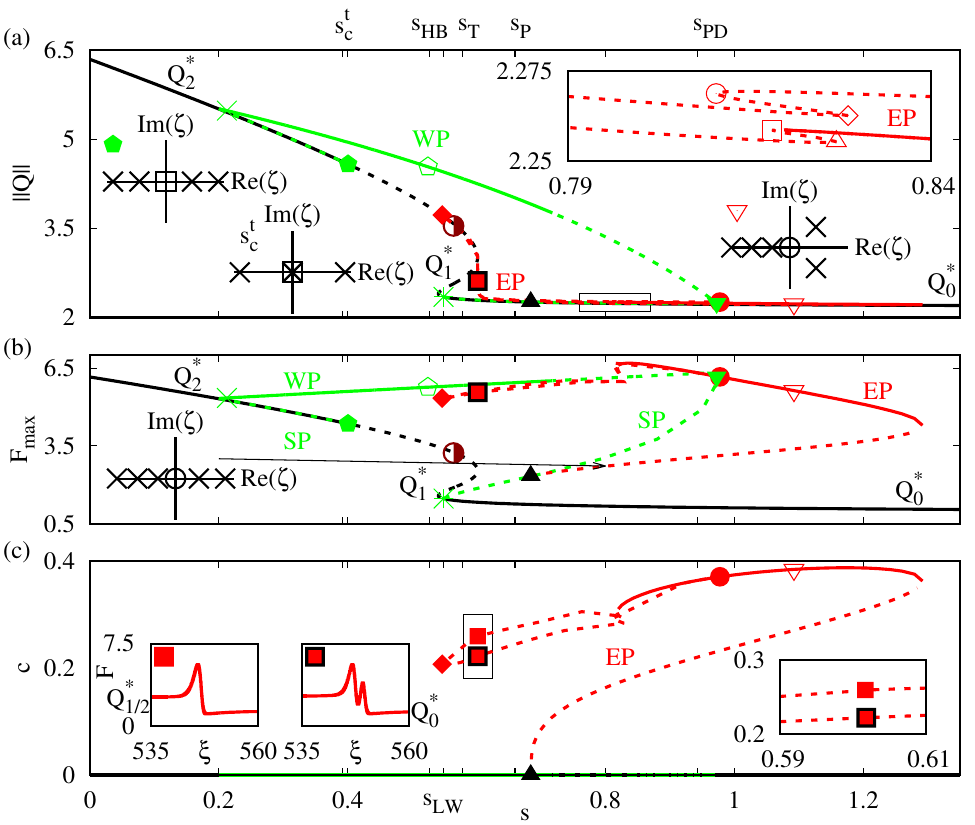}
    \caption{Bifurcation diagrams computed via numerical continuation as a function of $s$, showing excitable pulses (EPs), wave-pinning (WP) solutions, homogeneous steady states {(HSSs)} $Q^\ast_{0,1,2}${, and the stationary pulses (SP) that emerge from the long wavelength (LW) bifurcations} for $M=2$ and $L=1000\gg \lambda\approx3.09$ (where $\lambda$ is the wavelength leading to instability of $Q_2^\ast$). The solution branches are projected with~\cref{eq:Sobolev norm} in (a), the maximum value of $F$ in (b), and the speed $c$ in (c); solid (dashed) lines denote linearly stable (unstable) solutions. 
    The insets in (a) show the locations of several folds along the EP branch that are invisible otherwise, {and schematics of the spatial eigenvalues $\zeta$ of $Q_2^\ast$ at ($s=s_c^t$) and near (pentagon) the LW onset along the upper SP branch. The square denotes two zero eigenvalues due to mass conservation and spatial reversibility. This also holds for the LW of $Q_0^\ast$. The right inset in (a) shows the eigenvalue configuration of $Q_0^\ast$ at a stable EP (i.e., with respective speed $c$, inverted hollow triangle), where the circle shows the zero eigenvalue from mass conservation. The inset in (b) shows the eigenvalue configuration for EPs after the parity-breaking bifurcation. The insets in (c) show solutions at the T-points zoomed into the peaks and the separation of the two T-points.} The critical values $(s_c^t,s_{LW},s_{HB},s_T,s_P,s_{PD})=(0.409,0.548,0.565,0.601,0.684,0.978)$ denote the codimension-2 LW/finite wavenumber Hopf bifurcation (WB), the LW of $Q_0^\ast$, the homogeneous Hopf (HB), the T-point bifurcation (red square), the parity-breaking bifurcation of {SPs} (black triangle), and wavelength-doubling (red circle) bifurcations of EPs. The other symbols (pentagon, diamond, etc.)\ along the branches represent selected locations of solutions given in~\cref{fig:EP codim2 solutions}. The right semicircle denotes the location of the HB (see~\cref{fig:codim2 Hopf}). Other parameter values as in~\Cref{tab:par values} with $b=b_c\approx0.067$.} \label{fig:EP codim2 bif}
\end{figure}

\begin{figure}[tp]
    \centering
    \includegraphics[width=0.95\textwidth]{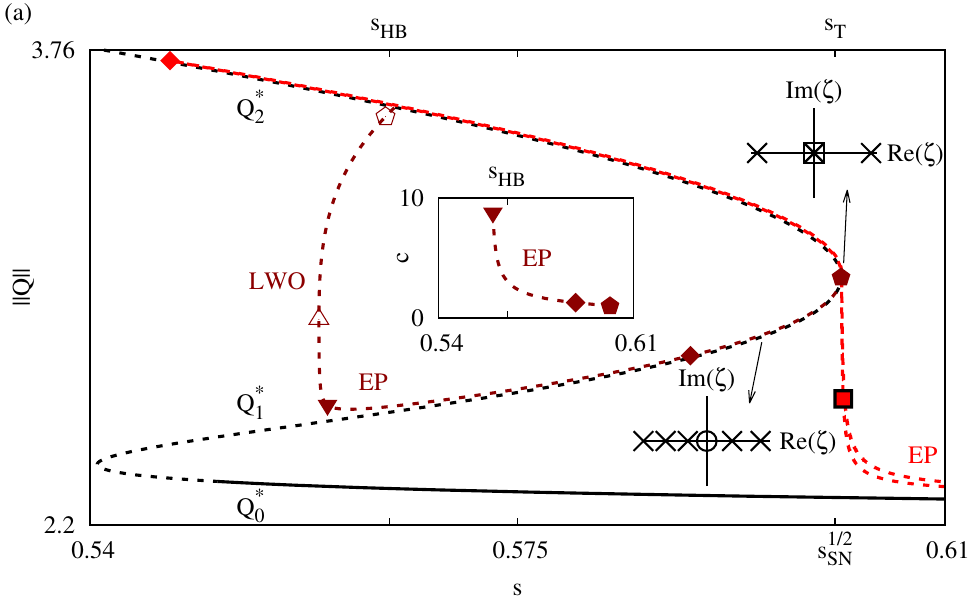}\\
    \includegraphics[width=0.92\textwidth]{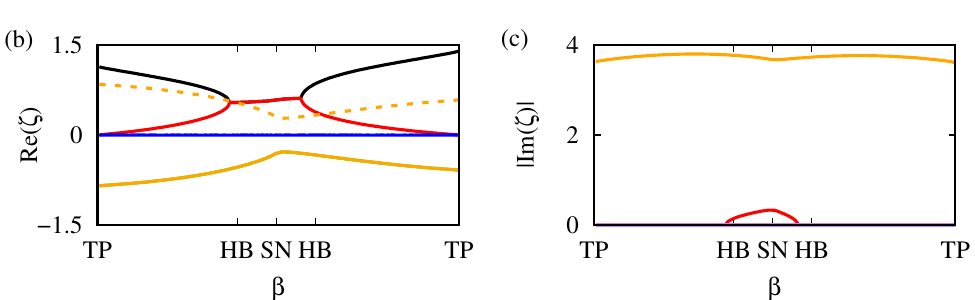}
    \caption{{(a) Bifurcation diagram computed via numerical continuation as a function of $s$, showing excitable pulses (EPs), homogeneous steady states $Q^\ast_{0,1,2}$, and the long wavelength oscillatory (LWO) solutions that emerge from the homogeneous Hopf bifurcation (HB) for $M=2$ and $L=2000$. The solution branches are projected with~\Cref{eq:Sobolev norm}. Solid (dashed) lines denote linearly stable (unstable) solutions. The left inset shows that the propagation speed $c$ of the EPs tends to zero as it approaches the saddle-node $SN_{Q_{1/2}^\ast}$. The right insets show schematics of the spatial eigenvalues $\zeta$ at the onset near the saddle-node $SN_{Q_{1/2}^\ast}$ and shortly after where $Q_1^\ast$ is the background state along the crimson red EP branch. The square denotes the two zero eigenvalues due to mass conservation and spatial reversibility when $c=0$ and the circle shows the persistent zero eigenvalue from mass conservation once spatial reversibility is lost. The critical values $(s_{HB},s_{SN}^{1/2},s_T)=(0.565,0.601,0.601)$ represent the homogeneous Hopf (HB), saddle-node of $Q_{1/2}^\ast$, and the T-point, respectively. (b) Real and (c) magnitude of the imaginary parts of the spatial eigenvalues $\zeta$ of the background state $Q_2^\ast$ along the red branch of EP solutions given between the T-points (TP, square) and the far-left fold (SN, diamond). Here, $\beta$ denotes the arc length along the red EP branch. The sixth eigenvalue is not shown but is negative, with $|\zeta|\sim\mathcal{O}(100)$. The dashed line in (b) denotes the magnitude of the real part of the negative real part complex conjugate pair (orange). The starting TP is given by red square in~\cref{fig:EP codim2 bif} and the ending TP by the bordered square. The other symbols (pentagon, diamond, etc.)\ along the branches represent selected locations of solutions given in~\cref{fig:EP codim2 solutions}d,e. Other parameter values as in~\Cref{tab:par values} with $b=b_c\approx 0.067$.}}\label{fig:codim2 Hopf}
\end{figure}

\begin{figure}[tp]
    \centering
    \includegraphics[width=\textwidth]{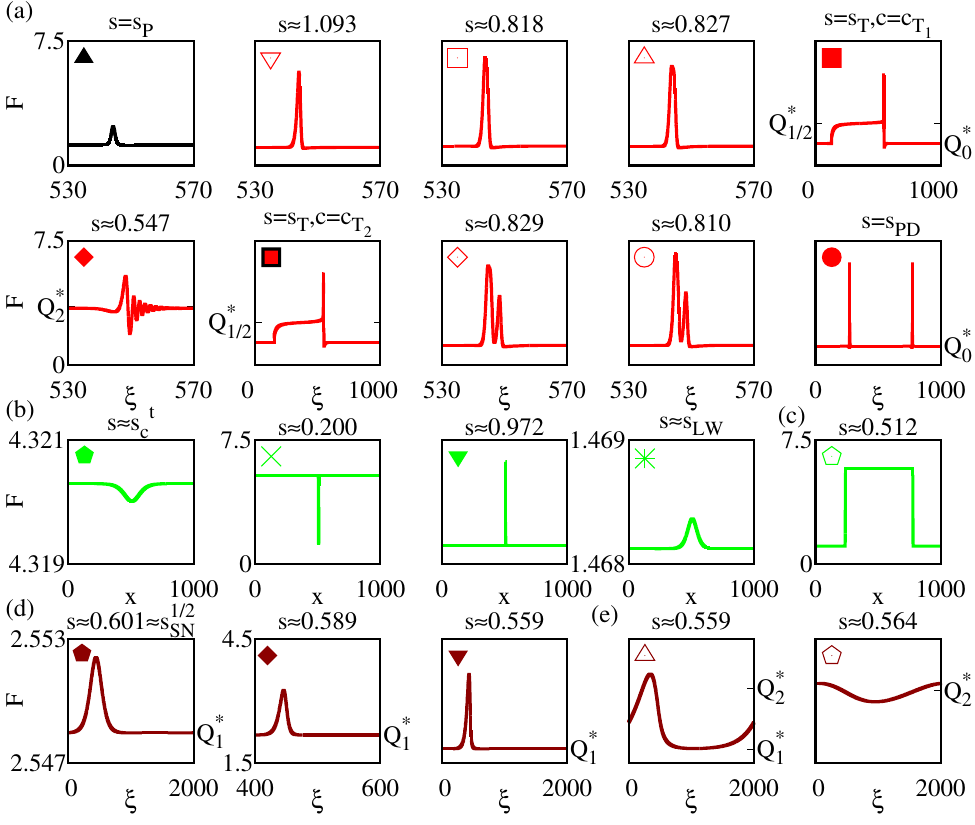}\\
    \includegraphics[width=\textwidth]{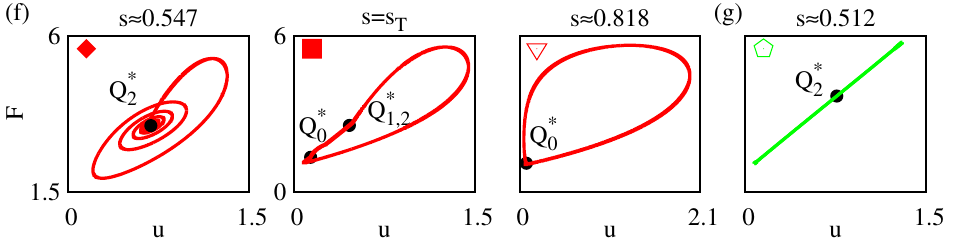}
    \caption{Solution profiles at selected locations along the (a) EP, (b) {SP}, and (c) WP branches in~\cref{fig:EP codim2 bif}, and the (d) EP and (e) {LWO branches in~\cref{fig:codim2 Hopf}}, where for visibility, the whole domain length of $L=1000$ is not always shown. Here $(s_{PD},s_P,s_T{, s_c^t,s_{LW}})\approx(0.978,0.684,0.601{,0.409,0.548})$ are defined as in~\cref{fig:EP codim2 bif}, $s_{SN}^{1/2}=0.601$ as in~\cref{fig:codim2 Hopf}, and $(c_{T_1},c_{T_2})\approx(0.260,0.222)$ denote the propagation speeds at the two T-points. Selected phase space projections are shown for EPs in (f) and WP solutions in (g), where in (g) only the homogeneous steady state $Q_2^\ast$ exists.}\label{fig:EP codim2 solutions}
\end{figure}

{Next, in~\Cref{fig:EP codim2 bif,fig:codim2 Hopf,fig:EP codim2 solutions}, we show that the bifurcation structure of coherent solutions for $M=2$ is qualitatively different from that of $M=4.5$. When $M=2$, EPs also bifurcate from SPs in a parity-breaking bifurcation at $s=s_P$ (triangle), but unlike EPs that are saddles (in space) for $M=4.5$, here, slightly after their onset, they become saddle-focus connections about $Q^\ast_0$. To show this, we compute the spatial eigenvalues $\zeta$, about the relevant HSS, here $Q^\ast_0$:
\begin{align}
    \begin{pmatrix}
        u\\
        v\\
        F\\
        U\\
        V\\
        f
    \end{pmatrix}-\begin{pmatrix}
        u^\ast\\
        v^\ast\\
        F^\ast\\
        0\\
        0\\
        0
    \end{pmatrix}\propto e^{\zeta\xi},
\end{align}}
{where the full system for which the computation is done, is being described in~\Cref{eq:PDE sys TW first}. The inset in~\cref{fig:EP codim2 bif}b shows the spatial eigenvalue configuration about $Q_0^\ast$ of the EP branch shortly after the parity-breaking bifurcation but before the onset of saddle-focus connections. Here we see that the zero eigenvalue from spatial reversibility becomes negative but initially all eigenvalues are real.} Afterwards, the solutions stabilize after the first fold and destabilize at the subsequent left fold (hollow square). {The inset in~\cref{fig:EP codim2 bif}a shows the spatial eigenvalue configuration about $Q_0^\ast$ of the EP branch (inverted hollow triangle), where we see that the trailing edge has become oscillatory.} After the EPs destabilize, the branch folds again before reaching a T-point bifurcation (square) at $s=s_T$ (see~\cref{fig:EP codim2 solutions}a). Here, the solutions form a homoclinic cycle with $Q_0^\ast$ at the saddle-node, i.e., the solutions briefly plateau at $Q_2^\ast=Q_1^\ast$ (see square in~\cref{fig:EP codim2 solutions}a). After passing the T-point, the spatially oscillatory tail becomes a leading edge decaying to $Q_2^\ast$, as shown by the profile at the left fold (diamond), see also the phase portrait projection in~\cref{fig:EP codim2 solutions}f. This switch is possible by additional annihilation of real eigenvalues and formation of a double focus before reaching the left fold. {In~\cref{fig:codim2 Hopf}b,c, we show the spatial eigenvalues, $\zeta$, corresponding to the above transitions. Here we see that the solutions are saddle-focus connections towards the T-points (squares) and focus-focus connections around the far-left fold (diamond). At the HB of the HSS $Q_2^\ast$, the magnitudes of the real parts of the leading eigenvalues swap and there is a brief regime where saddle-focus Shil-nikov homoclinics emerge~\cite{kuznetsov2004}. Then, the branch passes the T-point again, where an additional peak appears out of the leading edge oscillations. The branch associated with a bounded two-peak state goes through two more folds (hollow diamond and circle), after which the peaks start to separate until reaching equidistance, and thus, the branch terminates in a wavelength-doubling bifurcation at $s=s_{PD}$ (a consequence of finite size domain)}. We omit the EP branch with wavelength $L=500$ as the branch is effectively the same as with wavelength $L=1000$. Note that similar bifurcation structures related to peak doubling near T-point bifurcations have also been observed in an RD model but in the absence of mass conservation~\cite{knobloch2023front}.

{Next, we address the steady solution branch that bifurcates from $s=s_c^t$. As with high mass ($M=4.5$), the solutions emerge subcritically from $Q_2^\ast$ and $Q_0^\ast$ but these onsets can be instabilities when $M=2$, as shown in~\Cref{fig:2par bif 1par disper,fig:EP codim2 bif}. From both onsets and with both masses, the bifurcating solutions are spatially localized~\cite{bergmann2018active,otsuji2007mass} stationary pulses (SP) that rapidly grow in their amplitudes toward the respective folds~\cite{brauns2020phase}. From $Q_2^\ast$, these localizations are holes while from $Q_0^\ast$, these are peaks and the onset is close to $SN_{Q^\ast_{0/1}}$ (this persists for other values of $b$, see also~\cref{fig:2par bif HSS}a). Note that similar behaviour is observed at saddle-nodes in non-gradient systems without mass conservation~\cite{burke2008classification,parra2016dark,yochelis2006reciprocal}. In the insets of~\cref{fig:EP codim2 bif}a, we show that the spatial eigenvalue configuration about $Q_2^\ast$ and $Q_0^\ast$ follows the typical configuration of eigenvalues required for localization, whereby a pair of double zero eigenvalues split into two reals, coexisting with an additional positive/negative real pair (i.e., intersection of two two-dimensional manifolds~\cite{knobloch2015spatial}).} After the folds, the hole and the peak widen forming WP solutions, where $s$ controls the width. Hence, in between the two folds, the solutions form a homoclinic-like cycle, resembling two locked fronts. However, both plateaus are not HSSs of the system, as shown for mass-conserved systems in~\cite{verschueren2017model} and seen by the profile and phase-portrait projection marked by the hollow pentagon in~\cref{fig:EP codim2 solutions}b,g. Stability computations show that the WP solutions are linearly stable from the left fold and up to $s\approx 0.718$, see~\Cref{fig:EP codim2 bif}.

{Finally and for completeness, we address the large-scale solutions that emerge from the HB when $M=2$, see~\Cref{fig:codim2 Hopf} for the resulting branch. Here, we compute the branch with domain length $L=2000$ since with $L=1000$ the structure of the branch is not fully captured. Near the HB, long wavelength oscillatory (LWO) solutions emerge with a wavelength that matches the domain size (circle). The solutions reach a fold and then become excitable pulses (EPs) bi-asymptotic to $Q_1^\ast$ at a homoclinic bifurcation (inverted triangle). The solutions then continue in $s$ and terminate as localized patterns (pentagon) at the saddle-node $SN_{Q^\ast_{1/2}}$. Note that propagating solutions can terminate at a saddle-node (SN) in~\cref{eq:model} since there is a double zero eigenvalue at $SN_{Q^\ast_{1/2}}$ by mass conservation and these eigenvalues immediately turn complex as the wavenumber $q$ increases. It is also important to note that the propagation speed tends to zero as the solutions approach the saddle-node $SN_{Q^\ast_{1/2}}$ as expected. The spatial eigenvalues, $\zeta$, along the EP branch near the saddle-node are shown in the insets of~\cref{fig:codim2 Hopf}a and are computed assuming that $c=0$ at $SN_{Q^\ast_{1/2}}$. At $SN_{Q^\ast_{1/2}}$, there are four zero eigenvalues, two from the saddle-node and two from mass conservation with spatial reversibility. Immediately after $SN_{Q^\ast_{1/2}}$, the two zero eigenvalues from the saddle-node split into reals on either side of the imaginary axis and the extra zero eigenvalue from spatial reversibility becomes negative. Due to the numerical inaccuracies, it is unclear in which order the zero eigenvalues split into the negative half-plane. Hence, the EPs that form along this branch are saddle homoclinic orbits. Notably, there are no T-points as with the $M=4.5$ EP branch, i.e., for $M=2$ the solutions bifurcate as LWO solutions and form into EPs bi-asymptotic to $Q_1^\ast$ at a homoclinic bifurcation, and do not transition into pulses about $Q_0^\ast$ as in the case of $M=4.5$.}

In summary, the analysis for large domains in this section reveals a nontrivial organization of coherent solutions depending on the mass $M$ of the $(u,v)$-subsystem. Our key result for large domains is that WP solutions are linearly stable for low mass ($M=2$). This result enables our investigation into the stable coexistence of WP solutions with travelling waves. We next consider domains of moderate size.

\begin{figure}[tp]
    \centering
    \includegraphics[width=\textwidth]{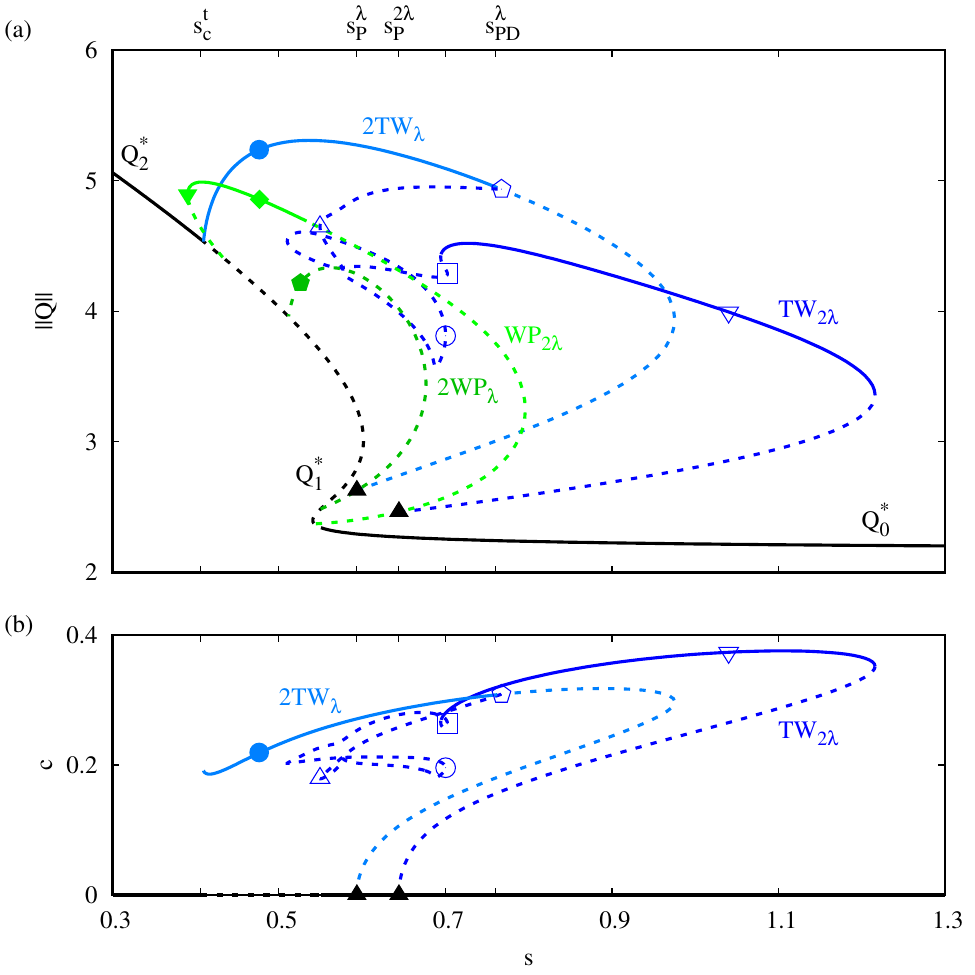}
    \caption{Bifurcation diagrams computed via numerical continuation as a function of $s$, showing the homogeneous steady states $Q^\ast_{0,1,2}$, and travelling waves (TWs) and wave-pinning (WP) solutions with wavelengths $\lambda$ and $2\lambda$ for $M=2$ and $L=2\lambda\approx2\cdot3.09$ (where $\lambda$ is the critical wavelength at the instability onset of $Q_2^\ast$). The solution branches are projected with the Sobolev norm~\cref{eq:Sobolev norm} in (a) and the propagation speed $c$ in (b); solid (dashed) lines denote linearly stable (unstable) solutions. The notation 2TW$_{\lambda}$ denotes two spatial copies of travelling waves with wavelength $\lambda$. Here, TWs and WP solutions form a bistable region with wavelengths $\lambda$ and $2\lambda$, respectively. The critical values $(s_c^t,s_P^\lambda,s_P^{2\lambda},s_{PD}^\lambda)=(0.409,0.593,0.664,0.760)$ denote the codimension-2 long wavelength (LW)/finite wavenumber Hopf (WB) onset, the parity-breaking bifurcations (black triangles) at wavelengths $\lambda$ and $2\lambda$, respectively, and the wavelength-doubling (hollow pentagon) bifurcation of $\lambda$. The symbols (circle, diamond, etc.)\ represent selected locations of solution profiles given in~\cref{fig:codim2 2L sols}. Other parameter values as in~\Cref{tab:par values} with $b=b_c\approx0.067$.} \label{fig:codim2 bif}
\end{figure}

\begin{figure}[tp]
    \centering
    \includegraphics[width=\textwidth]{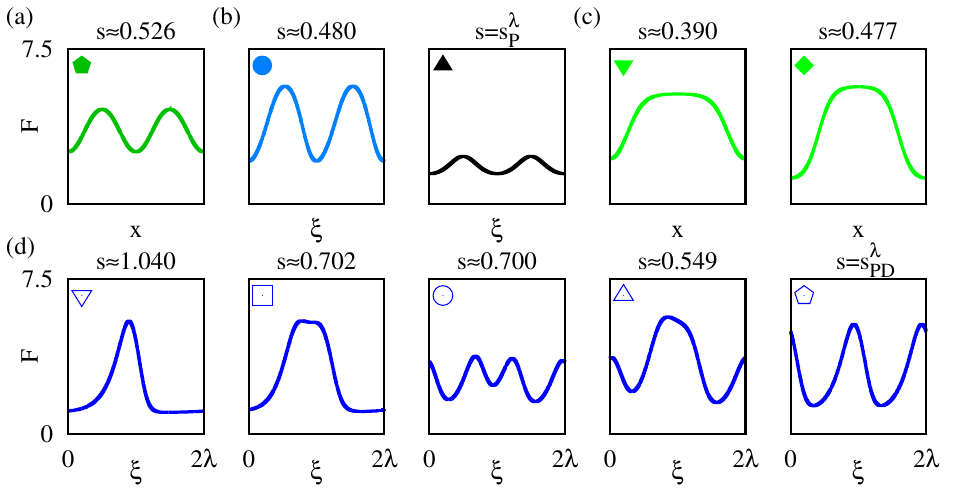}
    \caption{Solution profiles at selected locations in~\cref{fig:codim2 bif} along the branches (a) 2WP$_\lambda$ (b) 2TW$_\lambda$ (c) WP$_{2\lambda}$, and (d) TW$_{2\lambda}$. Here $(s_P^\lambda,s_{PD}^\lambda)\approx(0.593,0.760)$ are defined as in~\cref{fig:codim2 bif}.}\label{fig:codim2 2L sols}
\end{figure}

\begin{figure}[tp]
    \centering
    \includegraphics[width=\textwidth]{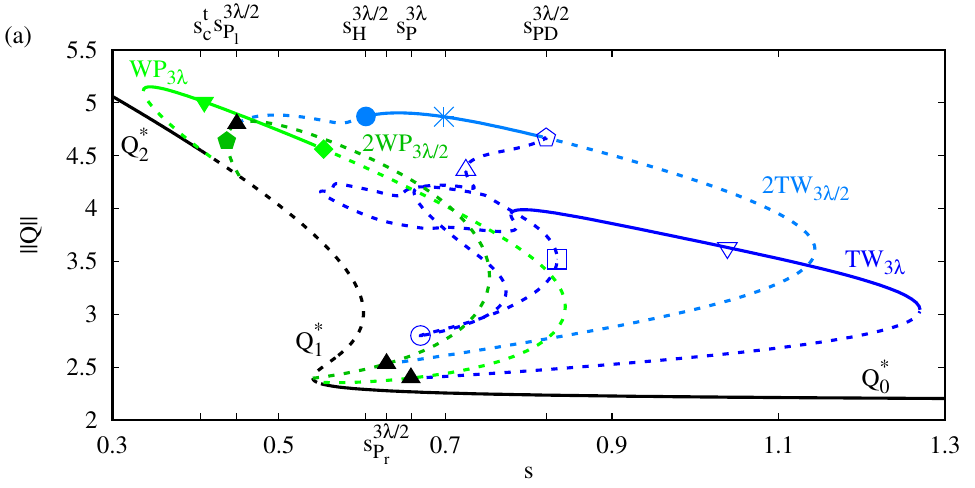}\\
    \includegraphics[width=\textwidth]{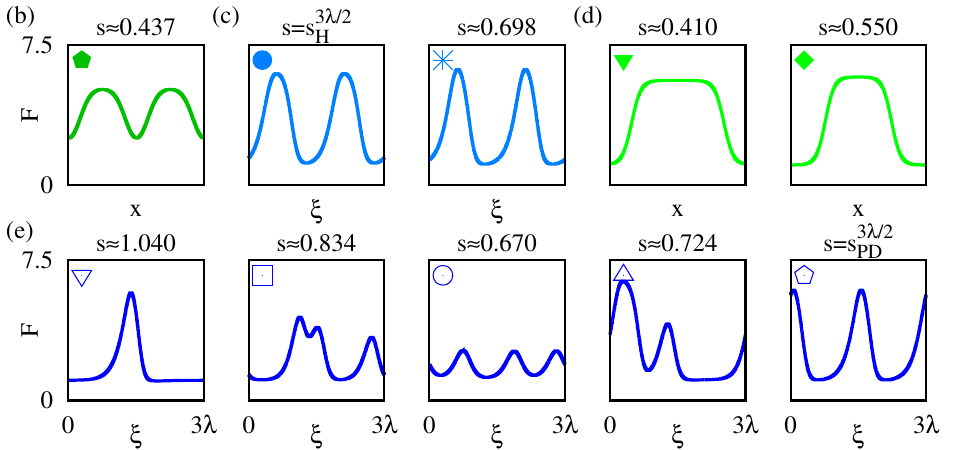}
    \caption{(a) Bifurcation diagram computed via numerical continuation as a function of $s$, showing the homogeneous steady states $Q^\ast_{0,1,2}$, and travelling waves (TWs) and wave-pinning (WP) solution branches with wavelengths $3\lambda/2$ and $3\lambda$ for $M=2$ and $L=3\lambda\approx3\cdot3.09$ (where $\lambda$ is the critical wavelength at the instability onset of $Q_2^\ast$). Here we only show projections with the Sobolev norm~\cref{eq:Sobolev norm}; solid (dashed) lines denote linearly stable (unstable) solutions. The notation 2TW$_{3\lambda/2}$ denotes two spatial copies of TWs with wavelength $3\lambda/2$, where the onsets of each one of the solution branches are at the respective parity-breaking bifurcations. The critical values are $(s_c^t,s_{P_l}^{3\lambda/2},s_{P_r}^{3\lambda/2},s_P^{3\lambda},s_{H}^{3\lambda/2},s_{PD}^{3\lambda/2})=(0.409,0.449,0.629,0.659,0.604,0.821)$, for the codimension-2 long wavelength (LW)/finite wavenumber Hopf (WB) bifurcation, the parity-breaking bifurcation at wavelengths $3\lambda/2$ on the left and right, and $3\lambda$, respectively, and the Hopf (H) and wavelength-doubling bifurcations of TW$_{3\lambda/2}$. (b-e) Selected solution profiles for (b) 2WP$_{3\lambda/2}$ (c) 2TW$_{3\lambda/2}$ (d) WP$_{3\lambda}$ and (e) TW$_{3\lambda}$ at corresponding points marked in (a). Other parameter values as in~\Cref{tab:par values} with $b=b_c\approx0.067$.} \label{fig:codim2 3L bif}
\end{figure}

\subsection{Coexistence and bistability for moderate domain size, $L\sim \mathcal{O}(m\lambda)$} \label{sec:low wavelength}

We now analyze a domain whose size is consistent with biological cells, namely much smaller domains than the large domain size in~\Cref{sec:other patterns full}. We keep the same parameter values as in~\Cref{fig:EP codim2 bif}. Since the primary bifurcation is the codimension-2 long wavelength/finite wavenumber Hopf instability (LW/WB), we choose the critical wavelength $\lambda \approx 3.09$ as a characteristic scale and the minimal wavelength of interest. We consider two cases, with domain lengths $L=2\lambda$ (\cref{fig:codim2 bif}) and $L=3\lambda$ (\cref{fig:codim2 3L bif}). In each case, we show WP and TW solutions with wavelength $L$ and $L/2$. Thus, the critical TW branch (TW$_{\lambda}$) is shown with domain length $L=2\lambda$. For all bifurcation diagrams on finite domains, we compute the stability of solutions for the largest domain length, unless otherwise stated. Therefore, solutions with smaller wavelengths are periodically extended to fit into the domain. This does not affect the curve of solutions but does change the linear stability onset along the branch. A notation of the form 2TW$_\lambda$ is used to indicate the travelling wave branch with wavelength $\lambda$ and 2 spatial copies.

Results for domain length $L=2\lambda$ are summarized in~\cref{fig:codim2 bif,fig:codim2 2L sols}. The 2WP$_\lambda$ solutions emerge for $s>s_c^t$ supercritically, due to finite size effects, and remain unstable until the branch terminates about $Q_1^\ast$, i.e., these solutions do not terminate at $Q_0^\ast$ like the WP solutions in large domains. The instability of WP solutions on small domains is unsurprising, cf.~\cite{mori2011asymptotic}. The 2TW$_\lambda$ branch is supercritical (as shown in~\Cref{sec:WnonlinA}), initially linearly stable, loses stability at $s=s_{PD}^\lambda$, and after a subsequent fold terminates at a parity-breaking bifurcation at $s=s_{P}^\lambda$. The TWs with wavelength $2\lambda$, however, are much different and resemble the structure of the EP branch. The solutions emerge from a parity-breaking bifurcation at $s=s_P^{2\lambda}$ of respective WP solutions, gain stability at the subsequent fold, and then lose stability at the next. Afterwards, the solutions transition to multiple peak solutions at the subsequent folds. In particular, we see that initially, the plateau of the peak expands (hollow square), then transitions into three-peak (hollow circle), and then into two-peak (hollow triangle) solutions but with different amplitudes and spacings. Finally, the branch reaches a wavelength-doubling bifurcation (as seen with EPs) at $s=s_{PD}^\lambda$, where the 2TW$_\lambda$ branch loses stability. The WP$_{2\lambda}$ branch resembles the bifurcation structure obtained for WP solutions in large domains: WP solutions emerge from $Q_2^\ast$ towards decreased $s$ values, and after the left fold, they gain stability, before reaching a subsequent fold and terminating at $Q_0^\ast$. Consequently, on domain size $L=2\lambda$, WP solutions stably coexist with TWs of two copies.

Increasing the domain length to $L=3\lambda$ introduces distinct bifurcation structures for TWs and WP solutions as shown in~\cref{fig:codim2 3L bif}, but also has many similarities. First, the WP$_{3\lambda}$ branch has a similar structure to the WP$_{2\lambda}$ branch. However, the TW$_{3\lambda}$ branch, which also emerges from a parity-breaking bifurcation of the WP$_{3\lambda}$ branch, goes through slightly different transitions from stable TW$_{3\lambda}$ to the period-doubling bifurcation of TW$_{3\lambda/2}$ at $s=s_{PD}^{3\lambda/2}$. In particular, the transition to a three-peak solution (hollow circle) requires an additional fold (hollow square) and the solution at the final fold before the period-doubling bifurcation is more localized with a broad trough (hollow triangle), as seen with the EPs. The 2TW$_{3\lambda/2}$ branch does not emerge from $Q_2^\ast$. Instead, these TWs emerge from two parity-breaking bifurcations, both from the 2WP$_{3\lambda/2}$ branch (which is everywhere unstable). On the right, at $s=s_{P_r}^{3\lambda/2}$ the parity-breaking bifurcation is similar to the previous TW branches, including the stability change at $s=s_{PD}^{3\lambda/2}$. However, these TWs lose stability in a Hopf bifurcation and terminate on the left, at an additional parity-breaking bifurcation, $s=s_{P_l}^{3\lambda/2}$. When $L=3\lambda/2$, TW$_{3\lambda/2}$ remains stable until the parity-breaking bifurcation at $s=s_{P_l}^{3\lambda/2}$ and WP$_{3\lambda/2}$ is stable between the initial fold (as with other WP branches) and the parity-breaking bifurcation. Therefore, a continuous transition from stable WP to TW solutions is observed when $L=3\lambda/2$.

Consequently, the most important outcome here is that multiple copies of TWs form a bistable region with WP solutions for a broad range of parameters and domain lengths, even though the bifurcation structure may differ in each case.

\begin{figure}[tp!]
    \centering
    \includegraphics[width=\textwidth]{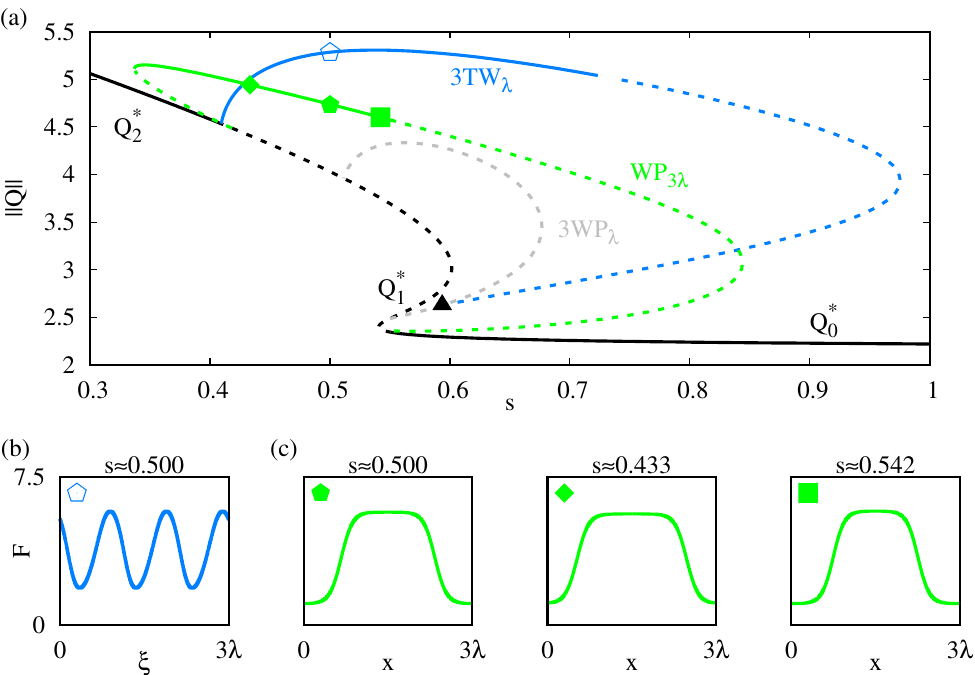}
    \caption{(a) Bifurcation diagrams computed via numerical continuation as a function of $s$, showing the homogeneous steady states $Q^\ast_{0,1,2}$, travelling waves (TWs) with wavelength $\lambda$, and wave-pinning (WP) solutions with wavelengths $\lambda$ and $3\lambda$ for $M=2$ and $L=3\lambda\approx3\cdot3.09$, where $\lambda$ is the wavelength at the instability onset of $Q_2^\ast$. The solution branches are projected with the Sobolev norm~\cref{eq:Sobolev norm}; solid (dashed) lines denote linearly stable (unstable) solutions. The notation 3TW$_{\lambda}$ denotes three spatial copies of TWs with wavelength $\lambda$. Note that the loss of stability for both WP$_{3\lambda}$ and TW$_{\lambda}$ is via Hopf bifurcations. The symbols (diamond, pentagon, square) denote locations of solutions used as components of initial conditions (ICs) in direct time-dependent simulations of~\cref{eq:model}. (b,c) Solution profiles in $F$ at locations indicated in (a), see also~\Cref{fig:basic sims}a-d for all solution components. Pentagons indicate ICs for~\cref{fig:convex kymo} while other symbols are used as ICs in~\cref{fig:perturbedWPcell}. Parameter values as in~\Cref{tab:par values} with $M=2$ and $b=b_c\approx0.067$.
    } \label{fig:sim bifs}
\end{figure}

\section{Pattern selection using time-dependent simulations} \label{sec:time simulations}

The bifurcation structures explored in~\Cref{sec:full bifurcation analysis} provide information on the types of solutions and their linear stability for various domain sizes $L$ and masses $M$ of the $(u,v)$-subsystem. These results do not yet inform us on the relative basins of attraction of the stable solutions, nor how they interact and compete. To explore these questions, we numerically integrate~\cref{eq:model} (for details see~\Cref{sec:simulation method}). We select initial conditions (ICs) that represent combinations of solution types and/or large amplitude perturbations: that is, we take combinations of stable WP and TW solutions with domain lengths $3\lambda$, i.e., picking each from the WP$_{3\lambda}$ and 3TW$_{\lambda}$ stable regime (see~\Cref{fig:sim bifs}). For each simulation, we only show the F-actin concentration, $F$; however, the active GTPase $u$ is also important in these perturbations as $u=0$ means there is locally no F-actin feedback to the GTPase. In~\Cref{sec:simulation method}, we show that the profiles of the other solution components are qualitatively similar and all non-negative.

\begin{figure}[tp!]
    \centering
    \includegraphics[width=\textwidth]{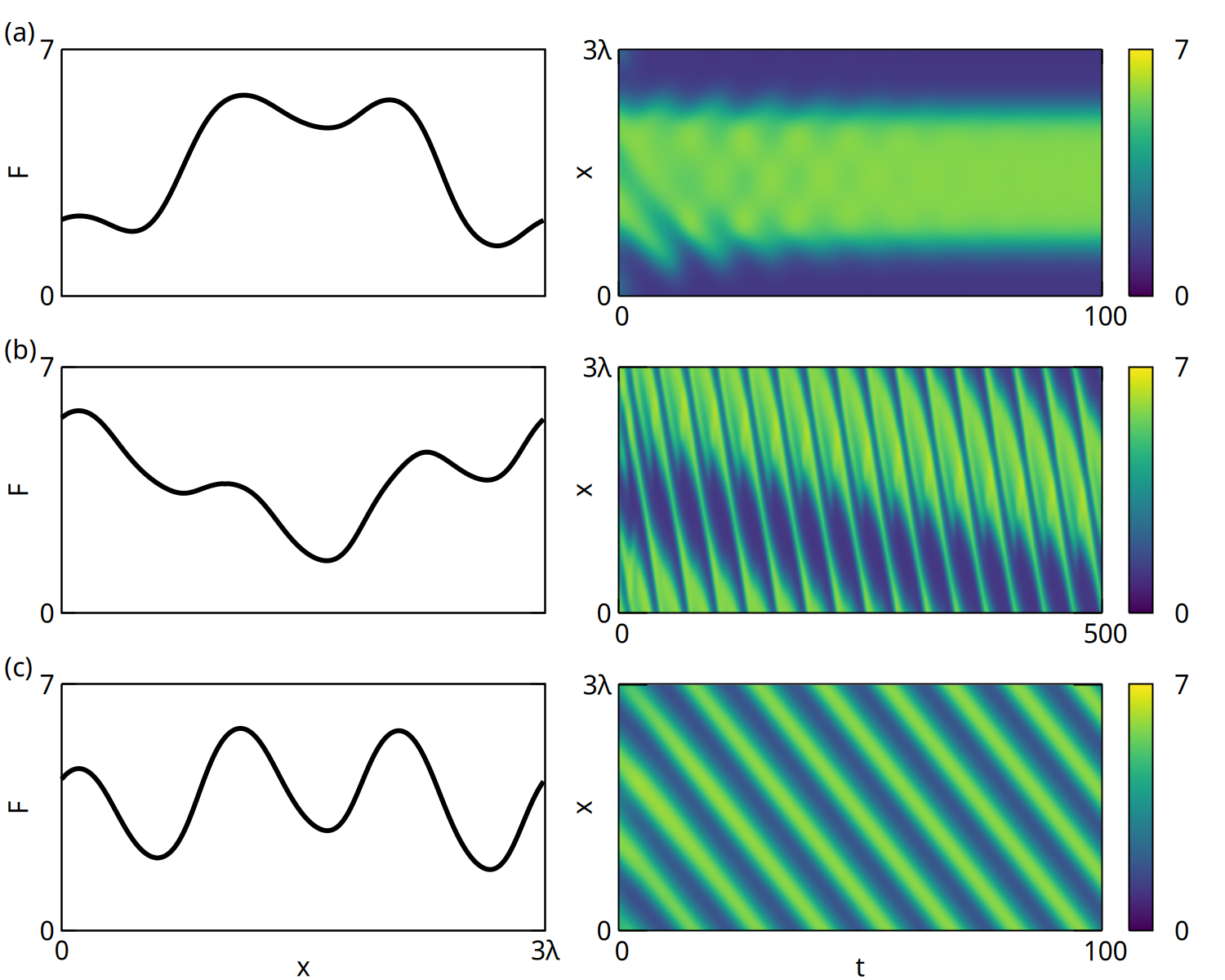}
    \caption{Direct numerical integration of~\cref{eq:model} with periodic boundary conditions, initialized with a convex combination~\eqref{eq:convex} of wave-pinning and travelling waves (left panels). Spacetime plots (``kymographs’’, right panels) showing $F$ as a heat map for (a) $\delta=0.25$, (b) $\delta=0.5$, and (c) $\delta=0.75$. In (b), we demonstrate a distinct kind of modulated solution that persists also at a much longer simulation time, $t=50000$; see also a supplementary movie\_Fig14b. Parameter values as in~\Cref{tab:par values} with $M=2, b=b_c\approx0.067$, $s\approx0.50$, and $L=3\lambda\approx3\cdot3.09$.} \label{fig:convex kymo}
\end{figure}

\subsection{Initial superposition of wave-pinning and travelling wave states, $L=3\lambda$}
\label{sec:simulations}
Setting $s=0.5$ (pentagons in~\Cref{fig:sim bifs}), we use ICs that superimpose the WP and TW solutions in the convex combination, 
\begin{align} \label{eq:convex}
Q_{IC}(x)=\delta Q_{TW}(x) + (1-\delta) Q_{WP}(x),
\end{align}
with $0\le \delta \le 1$ as a weighting parameter. \Cref{fig:convex kymo} shows ICs with $\delta=0.25,0.5,0.75$ (left) and space-time plots (a.k.a. kymographs, right) of the ensuing dynamics of $F$. As expected, the time-dependent simulation converges to the solution that is weighted more heavily in the IC. When the WP solution dominates ($\delta < 0.5$, \Cref{fig:convex kymo}a), oscillations induced by the TW linger for a while, but are eventually damped out. For $\delta>0.5$ (\Cref{fig:convex kymo}c), where TWs initially dominate, convergence to the pure TW solution is almost immediate. The weighting $\delta=0.5$ (\Cref{fig:convex kymo}b) leads to an exotic solution resembling a modulated travelling wave. However, one of the two peaks widens and splits, emanating an additional wave that dampens along the leading edge of the original peak (see supplementary movie\_Fig14b). This behaviour persists for a long time ($t\approx50000$), but we cannot assure it is an asymptotic solution. Therefore, complex time-dependent solutions and/or long transients are also possible.

\begin{figure}[tp!]
\centering
    \includegraphics[width=\textwidth]{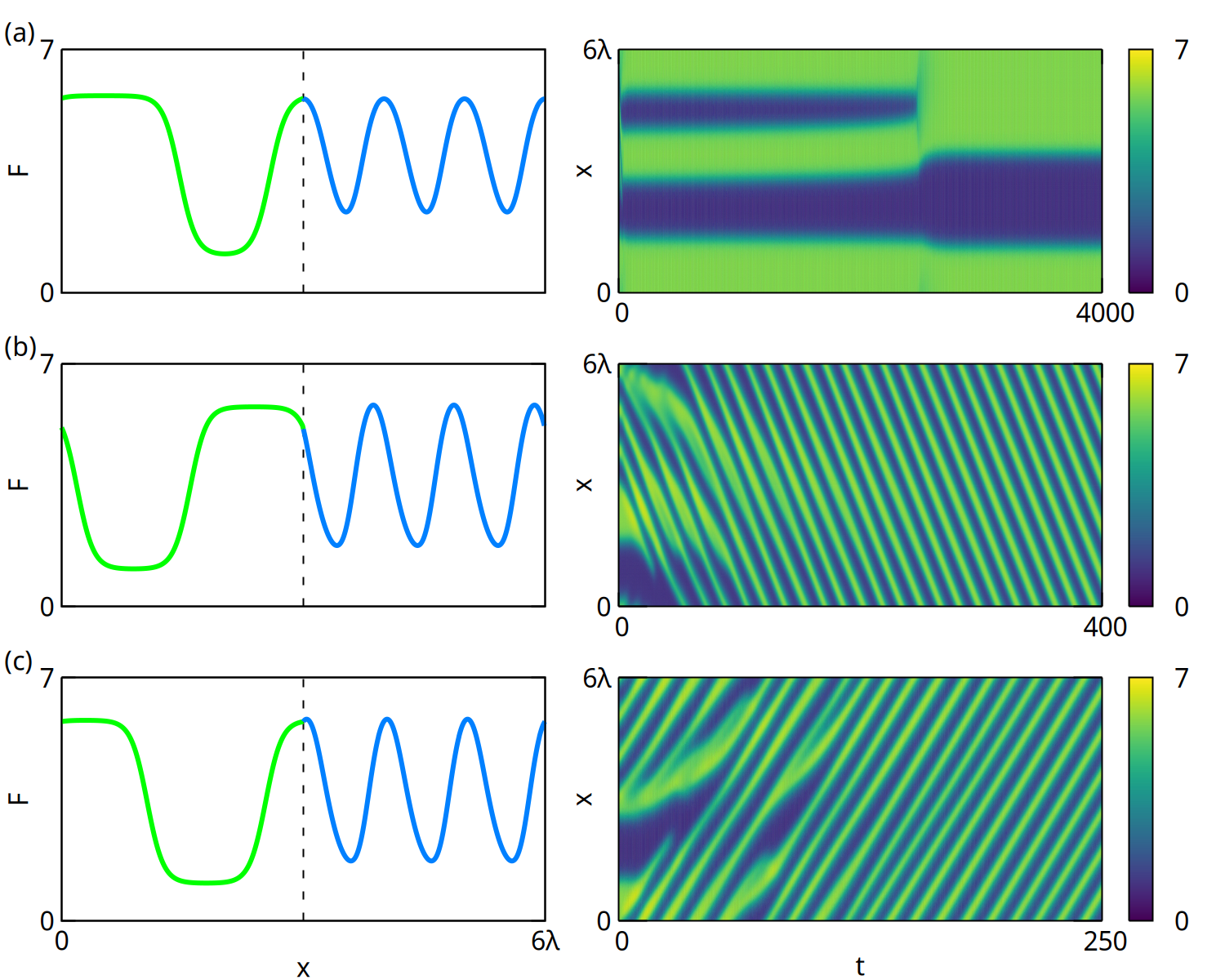}
    \caption{Direct numerical integration of~\cref{eq:model} with periodic boundary conditions, initialized with wave pinning (WP) solutions and 3-period travelling waves (TWs) of wavelength $\lambda$, on adjoining parts of the same domain (left panels). Spacetime plots (right panels) showing $F$ as a heat map for (a) $s\approx0.45$, (b) $s\approx0.52$, and (c) $s\approx0.53$. In (b) the asymptotic TWs are with a wavelength of $6\lambda/5$. Parameter values as in~\Cref{tab:par values} with $M=2, b=b_c\approx0.067$, and $L=6\lambda\approx6\cdot3.09$.} \label{fig:WP TW glue kymo 6}
\end{figure}

\begin{figure}[tp]
    \centering
    \includegraphics[width=\textwidth]{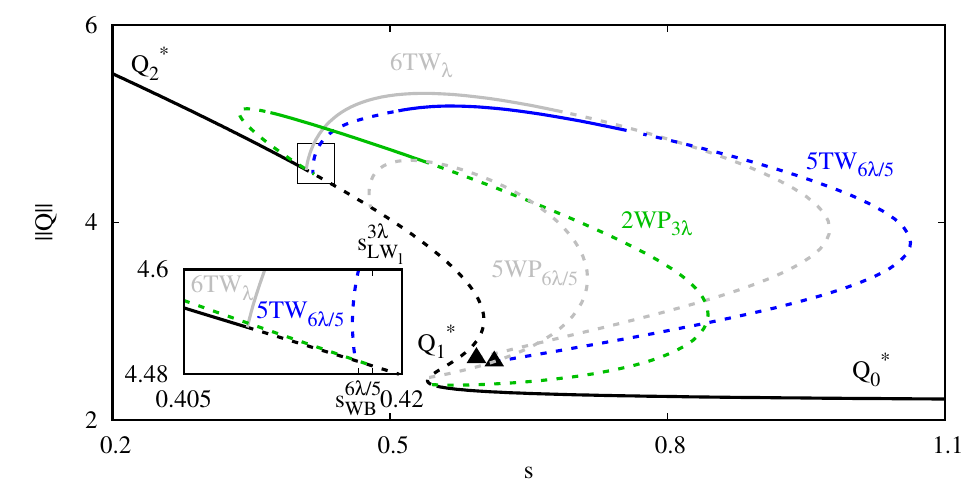}
    \caption{Bifurcation diagram computed via numerical continuation as a function of $s$, showing the homogeneous steady states $Q_{0,1,2}^\ast$ and the secondary solutions found in~\Cref{fig:WP TW glue kymo 6}a,b, with $L=6\lambda\approx6\cdot3.09$, where $\lambda$ is the wavelength at the instability onset of $Q_2^\ast$; solid (dashed) lines denote linearly stable (unstable) solutions. We showcase two-copies of WP$_{3\lambda}$ (2WP$_{3\lambda}$, green) and five-copies of TW$_{6\lambda/5}$ (5TW$_{6\lambda/5}$, blue). The inset shows a zoom into the bifurcation onsets, where both 2WP$_{3\lambda}$ and 5TW$_{6\lambda/5}$ are subcritical. The labels $(s_{WB}^{6\lambda/5},s_{LW_l}^{3\lambda})\approx(0.417,0.418)$ denote the $s$-values of the onset from $Q_2^\ast$ of the TW$_{6\lambda/5}$ and WP$_{3\lambda}$ branches, respectively. For comparison, we also plot the typical TW$_\lambda$ branch and the reference WP$_{6\lambda/5}$ (both in gray). Parameter values as in~\Cref{tab:par values} with $M=2$ and $b=b_c\approx0.067$.}  \label{fig:codim2 bif glue}
\end{figure}

\begin{figure}[tp]
    \centering
    \includegraphics[width=\linewidth]{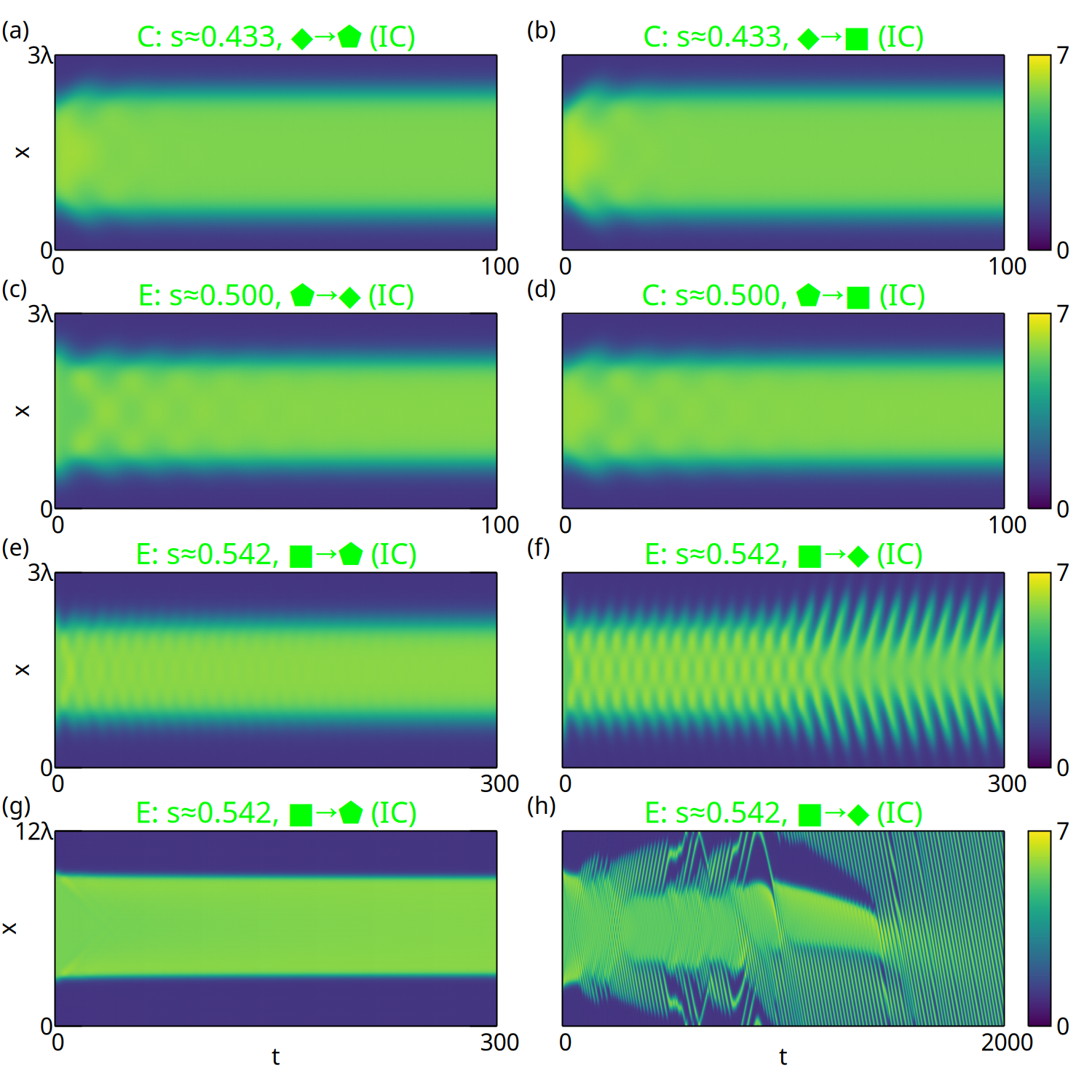}
    \caption{(a-h) Spacetime plots showing $F$ as a heat map obtained via direct numerical integration of~\cref{eq:model} with periodic boundary conditions, initialized with wave pinning (WP) solutions obtained from selected locations given by symbols in~\cref{fig:sim bifs}. The first letter in the titles denotes the type of numerical experiment: compression (C) and expansion (E) with respect to the steady-state WP solution at the given $s$ value. The symbols ${\color{green} \blacklozenge\to\blacksquare}$ in the titles correspond to selected locations in~\cref{fig:sim bifs}, where the parameter values are given by ${\color{green} \blacklozenge}$ and the initial condition is ${\color{green} \blacksquare}$. The profiles of each component of the initial conditions are given in~\cref{fig:basic sims}a-d of~\Cref{sec:simulation method}. Parameter values as in~\Cref{tab:par values} with $M=2$, $b=b_c\approx0.067$, and (a-f) $L=3\lambda\approx3\cdot3.09$ and (g,h) $L=12\lambda\approx 12\cdot3.09$, see also the supplementary movie\_Fig17f for (f).
    }
    \label{fig:perturbedWPcell}
\end{figure}

\subsection{Initial fronts of wave-pinning and travelling wave states, $L=2\cdot 3\lambda$}
\label{sec:simulationsInv}
Next, we consider interactions of TWs with WP solutions when they are initially in adjoining parts of the domain. We considered the effect of the parameter $s$ (which, recall, controls the width of the WP plateau) on the ability of TWs to invade and dominate. As shown in~\cref{fig:WP TW glue kymo 6}, initial conditions ($Q_{IC}$) combine a WP solution (on $0<x<3\lambda$) glued continuously to a TW (on $3\lambda<x<6\lambda$). We picked values of $s\approx 0.45, 0.52, 0.53$ (above $s_c^t$), according to:
\begin{align} \label{eq:WP TW glue IC}
    Q_{IC}(x)=\begin{cases}
        Q_{WP}(x) & x \leq 3\lambda\\
        \begin{pmatrix}
            u_{TW}(x-\rho_1)\\
            v_{TW}(x-\rho_2)\\
            F_{TW}(x-\rho_3)
        \end{pmatrix} & x>3\lambda
    \end{cases}
\end{align}
where $Q_{WP}$ and $Q_{TW}$ are pure WP and 3-period TWs with wavelength $\lambda$. The values of spatial shifts, $\rho_{1,2,3}$, are chosen so that $Q_{IC}$ are continuous. 

For $s=0.45$, TWs are rapidly damped, and a transient coarsening (of about 2500 time units)~\cite{kolokolnikov2006mesa} takes place before a single-plateau WP solution emerges (~\cref{fig:WP TW glue kymo 6}a). For $s=0.52$, we find that TWs rapidly take over (see~\Cref{fig:WP TW glue kymo 6}b); however, the asymptotic wavelength is $6\lambda/5$. For $s\approx0.53$, TWs also persist and the wavelength $\lambda$ is retrieved. We note that the choice of wavelengths depends on the initial condition (i.e.\ the shifting parameters $\rho_{1,2,3}$).

Motivated by numerical simulations in~\Cref{fig:WP TW glue kymo 6}a,b, we plot a bifurcation diagram (\cref{fig:codim2 bif glue}) showcasing the two-copies of WP$_{3\lambda}$ (green) and five-copies of secondary travelling waves of TW$_{6\lambda/5}$ (blue), respectively. As can be seen from~\cref{fig:codim2 bif glue}, 2WP$_{3\lambda}$ is stable at $s=0.45$ but the intermediate solution in~\cref{fig:WP TW glue kymo 6}a is a bimodal WP solution with differing wavelengths and is thus not two copies of a WP$_{3\lambda}$ solution. The TW solution with wavelength $6\lambda/5$ is one of the secondary TWs bifurcating for $s>s_c^t$ (\Cref{fig:codim2 bif glue}). For completeness, we also plot the WP$_{6\lambda/5}$ branch, noting that it emerges from $Q_1^\ast$ (as does the WP$_{\lambda}$ branch). The branch TW$_{6\lambda/5}$, however, has notable differences from the critical branch (TW$_{\lambda}$): it is subcritical rather than supercritical and does not gain stability at the initial fold (inset,~\Cref{fig:codim2 bif glue}). This implies that the stability region is entirely contained on the upper branch, bounded away from the folds. This follows the expected behaviour of TWs that emerge from secondary bifurcations (see supplementary materials of~\cite{Yochelis2022}). We also see that TWs with wavelength $6\lambda/5$ follow the primary branch of bifurcating TWs (TW$_\lambda$), meaning the branch connects the secondary finite wavenumber Hopf and the parity-breaking bifurcations (see~\Cref{fig:codim2 bif glue}).

\subsection{Robustness of wave-pinning states} \label{sec:biological significance}

A key characteristic of a WP solution is the plateau width, corresponding to F-actin. From~\Cref{sec:full bifurcation analysis}, we know that the plateau width of the WP solutions is controlled by $s$. So here, we address the role of large (nonlinear) perturbations of WP solutions following changes in the width of the plateau by employing ICs representing compression or expansion of the WP plateau. For ICs, we use the parameter values and corresponding WP solutions at the diamond, pentagon, and square on the stable portion of the WP$_{3\lambda}$ branch in~\cref{fig:sim bifs} and simulate all possible permutations of distinct parameter sets. This construction preserves the value of $M$ (and all other parameters).

\Cref{fig:perturbedWPcell}a-f showcases the six simulations generated from all possible permutations of initial conditions and parameter sets with $L=3\lambda$. Note that the square represents a solution close to the onset of the WP$_{3\lambda}$ Hopf instability.~\Cref{fig:perturbedWPcell}a-d shows that WP solutions persist for $s$ values away from the instability (diamond and pentagon). Note that this behaviour persists for initial conditions along the entire stable portion of the WP branch (i.e., to the left fold of the WP$_{3\lambda}$ branch in~\cref{fig:sim bifs}). However, \cref{fig:perturbedWPcell}f shows that near the instability (square) the WP solution is lost with a large enough perturbation. The emergent behaviour is initially of small amplitude standing waves (SWs), which at later times ($t\gtrsim 150$) destabilize and become modulated TWs. This behaviour was verified but not shown with slightly smaller $s$ values and ICs to the right of the square in~\cref{fig:sim bifs}. We repeated the experiments (e,f) but with a domain length of $L=12\lambda$, as shown in (g,h); note that for $L=12\lambda$ the instability of WPs is at $s\approx0.574$. Integration shows a similarity between (e,f) and (g,h), although in (h) we observe richer dynamics than in (f), which is expected in increased domain sizes. Since the Hopf instability onset of WP solutions in $L=12\lambda$ is at higher $s$ values, we attribute the instability to the nature of the perturbations (possibly due to subcriticality of the Hopf bifurcation) resulting from the dynamics of leading eigenfunctions near the Hopf onset rather than from the proximity to the onset location. The complex conjugated eigenfunctions (see~\Cref{fig:basic sims}e,f) indeed show pronounced contributions at the WP top plateau region, where SW-like dynamics are observed. Hence, these numerical results suggest that WP solutions increase their robustness if they are closer to the left fold, i.e., wider top plateaus.

\section{Conclusions} \label{sec:discussion}

We analyzed a mass-conserving reaction-diffusion (RD) model~\cref{eq:model} on a 1D domain with (mostly) periodic boundary conditions mimicking a ring of the actin cortex at the edge of a cell. Bifurcation analyses (see~\Cref{sec:LSA,sec:WnonlinA,sec:full bifurcation analysis}) reveal a rich structure of possible solutions depending not only on control parameters but also on the total mass $M$ of the $(u,v)$-subsystem. We prescribe the conditions in which steady wave-pinning (WP) solutions (a.k.a.\ cell polarization) can stabilize and coexist with stable travelling waves (TWs). Specifically, we study the bifurcations about a codimension-2 long wavelength (LW) and finite wavenumber Hopf (WB) instability and show how these depend on the domain length{, i.e., unfolding via domain length, so that} the main results qualitatively persist and are not subject only to the codimension-2 LW/WB instability. 

By studying solutions with distinct spatial scales, we showcase the important role of domain size in selecting the kinds of patterns that can emerge. In large domains (see~\cref{fig:EP codim2 bif,fig:EP M45 bif}), we showed that the stability regions of WP solutions and excitable pulses (EPs) are well separated in parameter space (see~\cref{fig:EP codim2 bif}), regardless of $M$ values. This separation is generic, since WP solutions with stable regions primarily bifurcate subcritically in a decreasing $s$ direction, whereas EPs bifurcate in a parity-breaking bifurcation towards increasing $s$. (This was confirmed, but not shown in other parameter regimes.) The fact that both EPs and WP solutions stabilize only after the folds creates an inherent separation regime. Moreover, the bistability regime of EPs and WP solutions also depends on the location of the homogeneous Hopf bifurcation (HB), which is a precursor for the organization of EPs around a T-point. In the parameter set chosen here, the HB onset is far from the primary instability onset $s=s_c^t$, so the bistability region is absent. However, we cannot exclude bistability under a different parameter setting. This is left for future explorations. In the case of large mass $M$ (\cref{fig:EP M45 bif,fig:Arik bif}), the LW onset is ``enclosed'' by the HB balloon (\Cref{fig:2par bif HSS}), so that the WP solutions also remain unstable after the fold, automatically excluding bistability. Therefore, we believe that bistability between EPs and WP solutions is unlikely to be observed, although we do not definitively exclude this possibility. In the moderate domain size limit (\cref{fig:codim2 bif,fig:codim2 3L bif}), TW and WP solutions show bistability over a broad range of conditions since they are connected through the codimension-2 LW/WB bifurcation, regardless of the criticality of the TWs. 

{The effect of mass conservation appears more profound than expected~\cite{beta2020large} (and the references therein), near local or global bifurcations. For example, the former could relate to bifurcation criticalities and spatial modulations (e.g., see~\Cref{fig:perturbedWPcell}) in the unfolding of a codimension-2 LW/WB instability (see~\Cref{fig:codim2 wave onset}) and also more generally in the study of a codimension-3 bifurcation, where LW, WB, and HB instabilities occur simultaneously. This construction should exist based on the results shown in~\Cref{fig:2par bif HSS} and should extend the understanding of spatial localizations studied in the presence of HBs~\cite{fauve1990solitary,thual1988localized}. Further studies of global bifurcations under mass-conservation could relate to the organizing centers by T-points~\cite{knobloch2023front} and their connection to modulated waves~\cite{knobloch11986degenerate}, including frequency locking for fronts exhibiting spatial oscillations (see~\Cref{fig:EP M45 solutions}) and the possible existence of the Shil'nikov/WB bifurcation~\cite{hirschberg1993vsil,yochelis2015origin}.} Other avenues to further study mass-conserved RD systems and their implications for cell behaviours include the exploration of parameter space to acquire elaborated intuition about the initial conditions and perturbations that determine the selection of cell motility modes, especially informative for experimental setups, in the spirit of~\cite{holmes2012regimes,miller2023generation,Rens_Edelstein-Keshet_2021,zmurchok2020shape} or the study of interactions leading to cell deformations, as inspired by~\cite{camley2017crawling,cao2019plasticity,dobereiner2006lateral,doubrovinski2011cell,maree2012cells,moreno2020modeling,taniguchi2013phase,vanderlei2011computational,veerman2021beyond}. Such interactions between domain geometry and reaction-diffusion dynamics promises to add new challenges and insights.

Consequently, the incorporation of mass conservation not only admits distinct and nontrivial pattern formation characteristics in RD systems {due to the coexistence of steady solutions emerging from the long wavelength bifurcation and oscillations emerging from a finite wavenumber Hopf, but also sheds a fresh light on the mechanisms of actin waves.} From~\Cref{fig:cellbehav}, we can draw an analogy between the solutions and the cell motility modes: the WP solutions represent polarization, the single wavelength TWs correspond to cell turning, and TWs with two or more wavelengths reflect a ruffling-like state, cf.~\cite{TeamKeshet2024,dobereiner2006lateral,giannone2004periodic,yang2019two}. Increasing the spatial dimension and/or coupling structural deformations of the cell due to mechanical or stochastic effects, may also lead to disordered motility~\cite{cao2019plasticity,dreher2014spiral,hladyshau2021spatiotemporal,liu2021,machacek2006morphodynamic,Michaud2022,Moldenhawer2022Switching,moreno2020modeling,Weiner2007}. This study is, therefore, only the tip of the iceberg, which we hope will stimulate further investigations.

\vskip 0.2in
\appendix

\section{{Linear onset of bifurcations along homogeneous steady states}} \label{app:slices}
\begin{figure}
    \centering
    \includegraphics[width=\textwidth]{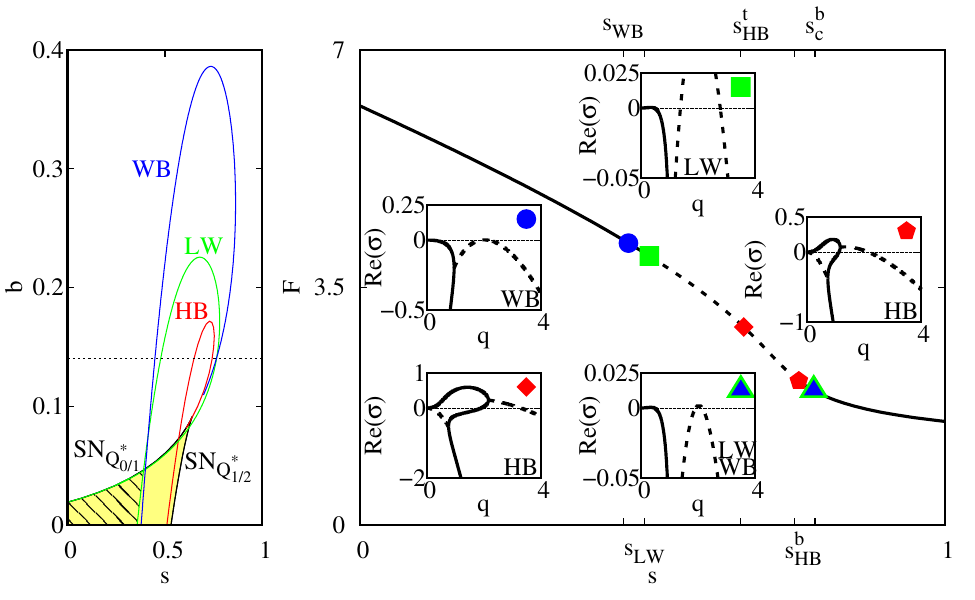}
    \caption{{(a) As in~\cref{fig:2par bif HSS}a but with a dotted line slice at $b\approx0.14$, where another codimension-2 long wavelength/finite wavenumber Hopf bifurcation emerges. (b) One-parameter bifurcation diagram for the F-actin component of the homogeneous steady states as a function of $s$, along the dotted line slice in (a). Solid (dashed) lines denote linear stability (instability). The values $(s_{WB},s_{LW},s_{HB}^t,s_{HB}^b,s_c^b)\approx (0.450,0.486,0.650,0.743,0.778)$ denote the finite wavenumber Hopf (WB) instability onset, long wavelength (LW) bifurcation, top and bottom Hopf bifurcations (HB), and the bottom codimension-2 LW/WB instability onset, respectively. Insets show dispersion relations at bifurcation onsets indicated by corresponding shapes (circle, square, etc.); the growth rate Re$(\sigma(q))$ of perturbations with wavenumber $q$ is a solid (dashed) line for real (complex conjugate) parts. Other parameter values as in~\Cref{tab:par values} with $M=2$.}}\label{fig:2 par appendix i}
\end{figure}

\begin{figure}
    \centering
    \includegraphics[width=\textwidth]{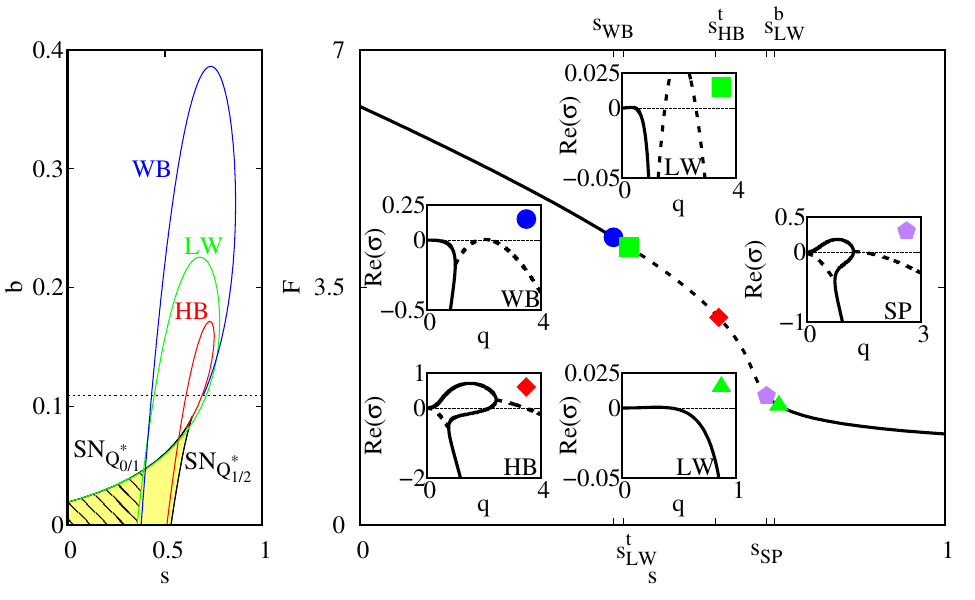}
    \caption{{(a) As in~\cref{fig:2par bif HSS}a but with a dotted line slice at $b\approx0.11$, where the locus of finite wavenumber Hopf bifurcations terminates. (b) One-parameter bifurcation diagram for the F-actin component of the homogeneous steady states as a function of $s$, along the dotted line slice in (a). Solid (dashed) lines denote linear stability (instability). The values $(s_{WB},s_{LW}^t,s_{HB}^t,s_{SP},s_{LW}^b)\approx (0.433,0.451,0.608,0.695,0.708)$ denote the finite wavenumber Hopf (WB) instability onset, top long wavelength (LW) bifurcation, top Hopf bifurcation (HB), the termination of finite wavenumber Hopf bifurcations (SP), and the bottom LW instability onset, respectively. Insets show dispersion relations at bifurcation onsets indicated by corresponding shapes (circle, square, etc.); the growth rate Re$(\sigma(q))$ of perturbations with wavenumber $q$ is a solid (dashed) line for real (complex conjugate) parts. Other parameter values as in~\Cref{tab:par values} with $M=2$.}}\label{fig:2 par appendix ii}
\end{figure}

\begin{figure}
    \centering
    \includegraphics[width=\textwidth]{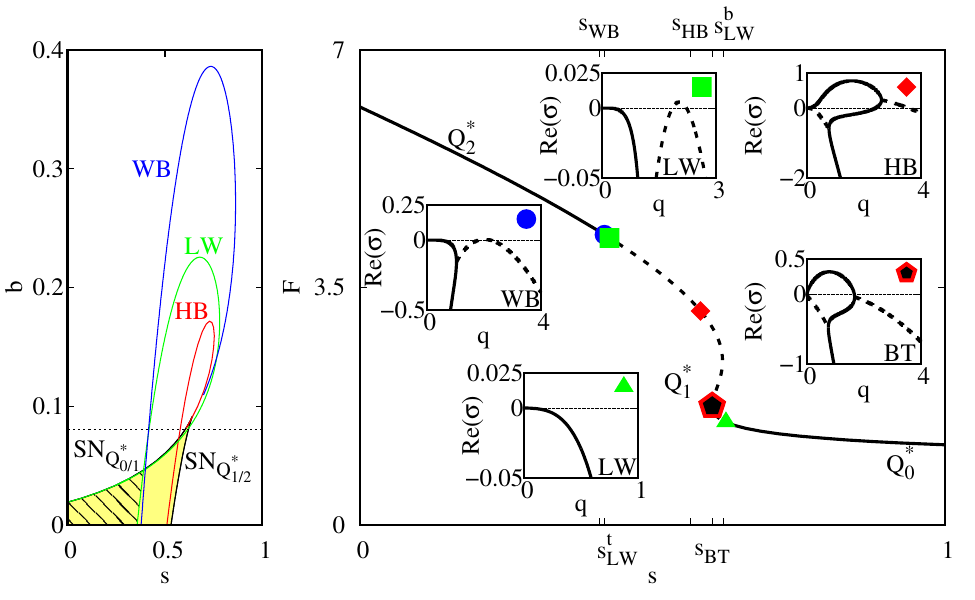}
    \caption{{(a) As in~\cref{fig:2par bif HSS}a but with a dotted line slice at $b\approx0.08$, where a codimension-2 Bogdanov-Takens bifurcation emerges. (b) One-parameter bifurcation diagram for the F-actin component of the homogeneous steady states as a function of $s$, along the dotted line slice in (a), where $Q^\ast_{0,1,2}$ are homogeneous steady states; solid (dashed) lines denote linear stability (instability). The values $(s_{WB},s_{LW}^t,s_{HB},s_{BT},s_{LW}^b)\approx (0.409,0.418,0.577,0.602,0.616)$ denote the finite wavenumber Hopf (WB) instability onset, top long wavelength (LW) bifurcation, top Hopf bifurcation (HB), Bogdanov-Takens (BT) bifurcation, and the bottom LW instability onset, respectively. Insets show dispersion relations at bifurcation onsets indicated by corresponding shapes (circle, square, etc.); the growth rate Re$(\sigma(q))$ of perturbations with wavenumber $q$ is a solid (dashed) line for real (complex conjugate) parts. Other parameter values as in~\Cref{tab:par values} with $M=2$.}}\label{fig:2 par appendix iii}
\end{figure}

\begin{figure}
    \centering
    \includegraphics[width=\textwidth]{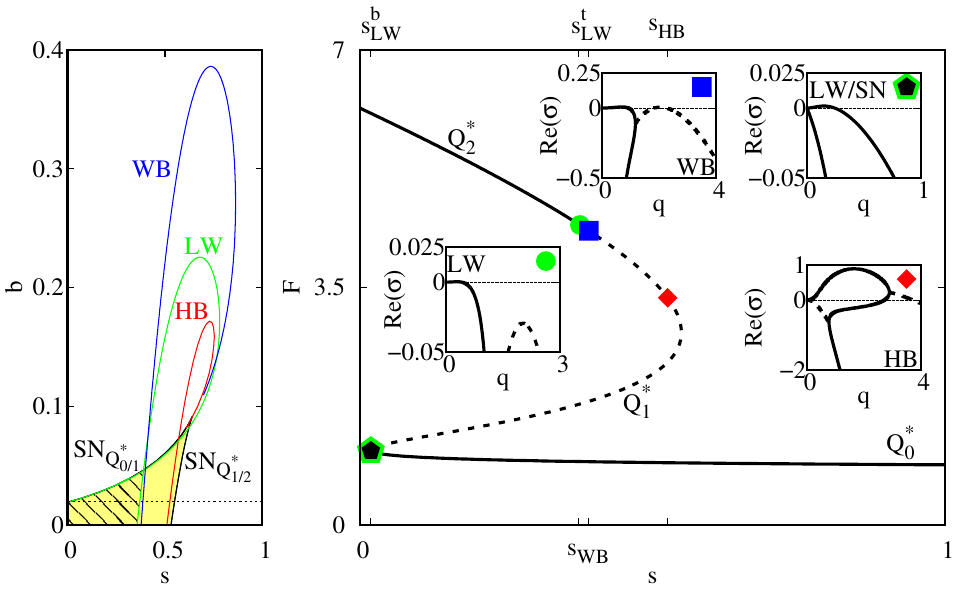}
    \caption{{(a) As in~\cref{fig:2par bif HSS}a but with a dotted line slice at $b\approx0.02$, where a codimension-2 long wavelength/saddle-node instability emerges. (b) One-parameter bifurcation diagram for the F-actin component of the homogeneous steady states as a function of $s$, along the dotted line slice in (a), where $Q^\ast_{0,1,2}$ are homogeneous steady states; solid (dashed) lines denote linear stability (instability). The values $(s_{LW}^t,s_{WB},s_{HB},s_{LW}^b)\approx (0.374,0.390,0.525,0.0182)$ denote the top long wavelength (LW) instability onset, finite wavenumber Hopf (WB) bifurcation, Hopf bifurcation (HB), and the bottom LW instability at the bottom saddle-node (SN), respectively. Insets show dispersion relations at bifurcation onsets indicated by corresponding shapes (circle, square, etc.); the growth rate Re$(\sigma(q))$ of perturbations with wavenumber $q$ is a solid (dashed) line for real (complex conjugate) parts. Other parameter values as in~\Cref{tab:par values} with $M=2$.}}\label{fig:2 par appendix v}
\end{figure}

{For completeness, we plot additional one-parameter bifurcation diagrams of homogeneous steady states along fixed $b$-value slices in~\cref{fig:2par bif HSS}a, where additional codimension-2 bifurcations emerge. \Cref{fig:2 par appendix i,fig:2 par appendix ii,fig:2 par appendix iii,fig:2 par appendix v} provide the bifurcation diagrams with dispersion relations at selected locations along the homogeneous steady state branches. In particular, we see another codimension-2 long wavelength (LW)/finite wavenumber Hopf (WB) instability when $b\approx0.14$ (triangle in~\cref{fig:2 par appendix i}b) and the termination of the WB locus for $b\approx0.11$. Here, there is a double zero eigenvalue for some $q=q_c$ and the slope of the dispersion relation is zero (see pentagon in~\cref{fig:2 par appendix ii}b). There is also a Bogdanov-Takens bifurcation when $b\approx0.08$ (pentagon in~\cref{fig:2 par appendix iii}b) and a codimension-2 long wavelength/saddle-node instability for $b\approx0.02$ (pentagon in~\cref{fig:2 par appendix v}b).}

\section{Bifurcation structure for the ventral actin wave model} \label{sec:Yochelis model}
Yochelis \textit{et al.}~\cite{Yochelis2022} considered a model describing actin dynamics on the ventral side of cells, comprising F-actin $N$, G-actin $S$, and an inhibitor $I$ affecting $N$, 
\begin{align}
    \begin{aligned} \label{eq:Yochelis model}
        \frac{\partial N}{\partial t}&=\frac{N^2S}{I+1}-N+D_N\frac{\partial^2N}{\partial x^2},\\
        \frac{\partial S}{\partial t}&=-\frac{N^2S}{I+1}+N+\frac{\partial^2S}{\partial x^2},\\
        \frac{\partial I}{\partial t}&=r_NN-r_II+D_I\frac{\partial^2I}{\partial x^2}.
    \end{aligned}
\end{align}
so that the total amount of actin is conserved,
\begin{align}
    M:=\frac{1}{L}\int_0^L[N(x,t)+S(x,t)]{\textrm d}x=\text{constant}.
\end{align}
Model details are given in~\cite{Yochelis2022}, where the F-actin and the G-actin correspond to our active and inactive GTPase and the inhibitor to F-actin. The main difference is that the inhibitor $I$ (our F-actin) does not promote depolymerization (inactivation of the GTPase) but instead inhibits polymerization of actin (activation of the GTPase).

\begin{figure}[tp]
    \centering
    \includegraphics[width=\textwidth]{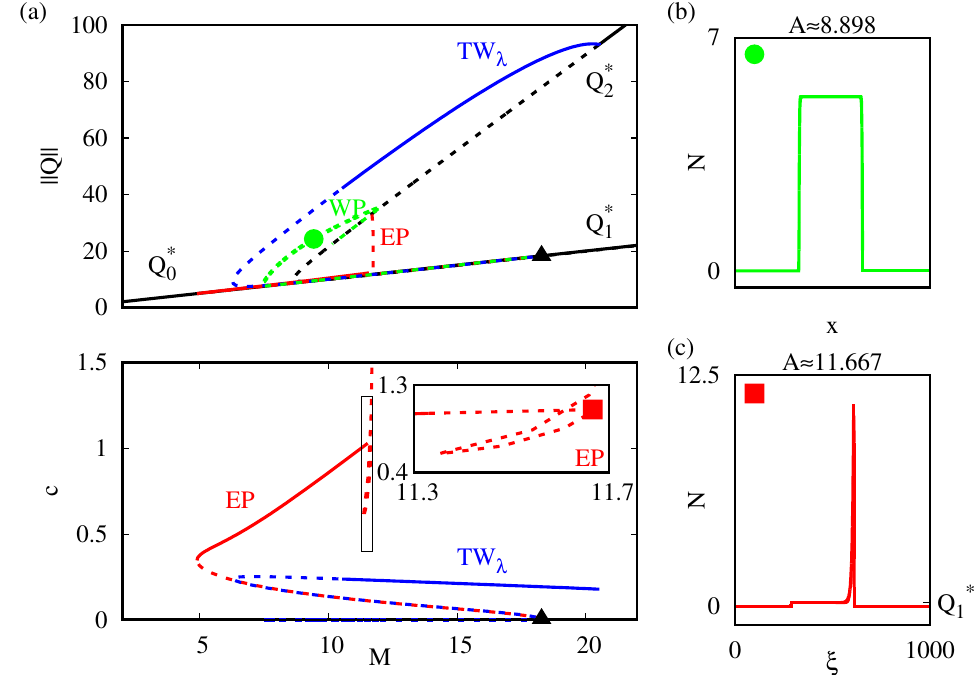}
    \caption{Bifurcation diagrams of~\cref{eq:Yochelis model} with respect to total mass $M$, showing the homogeneous steady states $Q^\ast_{0,1,2}$, travelling waves (TWs) with critical wavelength $\lambda$ (TW$_\lambda$), wave-pinning (WP) solutions, and excitable pulses (EPs) for $L=1000\gg\lambda\approx4.02$. The solution branches are projected with the Sobolev norm~\cref{eq:Sobolev norm} and the propagation speed $c$; solid (dashed) lines denote linearly stable (unstable) solutions. The inset in (a) shows the T-point bifurcation (square). In this setting, excitable pulses coexist with travelling waves while the wave-pinning solutions are all unstable. (b,c) Solutions showing a wave-pinning solution (circle) in (b) and a double front solution (square) at the T-point bifurcation (a heteroclinic cycle) in (c). Parameter values: $D_N=0.1$, $D_I=0.001$, $r_N=2$, and $r_I=0.3$.
    } \label{fig:Arik bif}
\end{figure}

\Cref{fig:Arik bif} shows the bifurcation structure and properties of travelling waves (TWs), excitable pulses (EPs), and wave-pinning (WP) solutions, along with the homogeneous steady states. (Note that wave-pinning solutions were not investigated in~\cite{Yochelis2022}.) In this setting, EPs and TWs form a bistability region while the stationary patterns are all unstable - a similar situation in the current study for $M=4.5$.

\section{Derivation of amplitude equations}\label{sec:amp_eq_app}
Here we discuss the main algebraic steps in the derivation of the amplitude equations~\eqref{eq:amps}. Substitution of~\cref{eq:firstorder}, {with  $s = s_{WB} + \epsilon^2 \widetilde s$ and $Q^\ast = Q_c^\ast(s_{WB}) + \epsilon^2 \widetilde Q^\ast$}, into~\cref{eq:model}, and separating into powers of $\epsilon$ (up to third order), yields at each $\mathcal{O}(\epsilon^n)$ the corresponding equation:
\begin{equation}\label{eq:orderi}
	\mathcal{\widetilde L}Q_n=\mathbi{R}_n,
\end{equation}
where $\mathcal{\widetilde L}:={\partial_t}  \mathbb{I} - \mathcal{L}(Q^\ast)|_{s = s_{WB}}${, $Q_n=(u_n,v_n,F_n)^T$,} and $\mathbi{R}_n$ includes nonlinear terms and depends on lower order contributions to the solution that are already known (meaning on $Q_1,...,Q_{n-1}$).

At $\mathcal{O}(\epsilon)$ we obtain the system 
\begin{equation}\label{eq:order_delta}
\mathcal{\widetilde L} Q_1=\mathbi{R}_1=\left( \begin{array}{c}
     0 \\
     0 \\
     0
    \end{array} \right), 
\end{equation}
which solutions give the so-called eigen-relations
\begin{subequations}\label{onsetrelation}
\begin{align}
B_{Lu} =& \frac{D_F q_c^2 + \theta + i \omega_c}{p_1 \theta} B_{LF} := b_{u} B_{LF},\\
B_{Lv} =&\dfrac{s_{WB} u_c^\ast}{b + \gamma{u_c^\ast}^2}B_{LF}\\
\nonumber &+\dfrac{1 + D_u q_c^2 + \bra{p_0 + p_1 u_c^\ast}s_{WB} + \bra{3 + 2 \gamma}{u_c^\ast}^2 - 2 \gamma M u_c^\ast + i \omega_c }{b + \gamma{u_c^\ast}^2} b_u B_{LF}:= b_{v} B_{LF},
\end{align}
\end{subequations}
where $(u_c^*,v_c^*,F_c^*)=(u^\ast,v^*,F^*)|_{s=s_{WB}}$.
Similarly, proceeding to the higher orders, we obtain:
\begin{equation}\label{eq:RHS2}
    \mathbi{R}_2=
    \BRA{\Bra{\bra{\gamma M - \bra{3 + \gamma} u_c^\ast} u_1 + 2 u_c^\ast \gamma v_1 - s_{WB} F_1} u_1}
    \left( \begin{array}{c}
     1 \\
     -1 \\
     0
    \end{array} \right),
\end{equation}
{\begin{multline}\label{eq:RHS3}
    \mathbi{R}_3 =
    - \frac{\partial}{\partial \tau} 
    \bra{ \begin{array}{c}
     u_1 \\
     v_1 \\
     F_1
    \end{array} }
    + \BRA{\Bra{\bra{2 \gamma M - 6 u_c^\ast} u_1  + 2 u^\ast_c \gamma \bra{v_1 - u_1} -F_1 s_{WB}} u_2}  
    \bra{ \begin{array}{c}
     1 \\
     -1 \\
     0
    \end{array} }\\
    + \{\Bra{2 \gamma u^\ast_c v_2 - s_{WB} F_2 + \bra{2 M \gamma - 6 u^\ast_c - p_1 s_{WB} - 4 u^\ast_c \gamma} \tilde u^\ast - (p_0 + 
    p_1 u^\ast_c) {\widetilde s}} u_1 
    \\ + 2 u^\ast_c \tilde u^\ast  v_1 \}
    \bra{ \begin{array}{c}
     1 \\
     -1 \\
     0
    \end{array} }
    + \BRA{\bra{\gamma v_1 - u_1}{u_1}^2 - (s_{WB} \tilde u^\ast + {\widetilde s} u^\ast_c) F_1}
    \bra{ \begin{array}{c}
     1 \\
     -1 \\
     0
    \end{array} }.
\end{multline}}
To avoid secular terms in $\mathbi{R}_n$, we use the \textit{Fredholm alternative} as a formal framework for applying the solvability condition by utilizing an inner product:
\begin{equation}\label{eq:innerprod}
	\langle f,g \rangle \equiv \frac{1}{T\lambda} \int_0^T\int_0^\lambda {\bar f} g\,{\text d}x {\text d}t,
\end{equation}
where $T=2\pi/\omega_c$ and $\lambda=2\pi/q_c$. This leads to a solvability condition
\begin{equation}\label{eq:solvability}
    \langle \mathbi{N},\mathbi{R}_n \rangle=\langle \mathbi{N},\mathcal{\widetilde L}\mathbi{Q}_n \rangle=\langle \mathcal{\widetilde L}^\dagger \mathbi{N},\mathbi{Q}_n \rangle = 0,
\end{equation}
where $\mathcal{\widetilde L}^\dagger$ is the adjoint of $\mathcal{\widetilde L}$ 
\\
\begin{equation*}
    \mathcal{\widetilde L}^\dagger = 
    \bra{\begin{array}{ccc}
        -\partial_{t} - D \partial^2_{x} + \varrho & -\varrho & -\theta p_1  \\
         -b -  \gamma u_c^{*2}  & -\partial_{t} - \partial^2_{x} + b + \gamma u_c^{*2} & 0\\
         s_{WB} u_c^* & -s_{WB} u_c^* & -\partial_{t} - D_F\partial^2_{x} + \theta
    \end{array}},
\end{equation*}
where $\varrho = 1 + p_0 s_{WB} + p_1 s_{WB} u_c^* + 3 u_c^{*2}+ 2  \gamma u_c^{*2} - 2 M u_c^* \gamma$,
and the components of the nullvector 
\[
\mathbi{N}=\left(\begin{array}{c}
    N_1\\
    N_2\\
    N_3
    \end{array}\right) e^{i\bra{\omega_c t\pm q_c x}}+c.c.
\]
are related by
\begin{subequations}
\begin{align}
   N_1 =& - \frac{\bigl(b + q_c^2 + \gamma {u_c^\ast}^2 - i \omega_c \bigr) \bigl(D_F q_c^2 + \theta - i \omega_c \bigr)}{s_{WB} u_c^\ast \bigl(q_c^2 - i \omega_c \bigr)} N_3,\\
   N_2 =& - \frac{\bigl(b + \gamma{u_c^\ast}^2 \bigr) \bigl(D_F q_c^2 + \theta - i \omega_c\bigr)}{s_{WB} u_c^\ast \bigl(q_c^2 - i \omega_c \bigr)} N_3.
\end{align}
\end{subequations}
{Applying the inner (scalar) product and integrating, we obtain that the solvability condition is automatically satisfied, leading to a solution
\begin{multline*}
    Q_2= \left(\begin{array}{c}
     a_1\\
     a_2\\
     a_3
    \end{array}\right)B_{LF}^2 e^{2 i \omega_c t + 2 i q_c x}
    + \left(\begin{array}{c}
    b_1\\
    b_2\\
    b_3
    \end{array}\right) B_{RF}^2 e^{2 i \omega_c t - 2 i q_c x}
    + \left(\begin{array}{c}
    c_1\\
    c_2\\
    c_3
    \end{array}\right) B_{LF} B_{RF} e^{2 i \omega_c t} \\
    +\left(\begin{array}{c}
    d_1\\
    d_2\\
    d_3
    \end{array}\right)  B_{LF} \overline{B}_{RF} e^{2 i q_c x}
    +\left(\begin{array}{c}
    h_1\\
    h_2\\
    h_3
    \end{array}\right) \bra{|B_{LF}|^2 + |B_{RF}|^2}
    + E \left(\begin{array}{c}
    u_1\\
    v_1\\
    F_1
    \end{array}\right)+ c.c.,
\end{multline*}
where the last term is a solution of the homogeneous problem \eqref{eq:order_delta}, $E$ is an arbitrary complex-valued constant that is set to zero, and the other} coefficients are obtained numerically using \textit{Mathematica} {by solving~\ref{eq:orderi} for respective terms at order $n=2$}.
Finally, inserting $Q_2$ into the form of $\mathbi{R}_3$, and expanding in powers of $e^{i\left( \pm  \omega_c t \pm  q_c x \right)}$, yields after applying~\cref{eq:solvability} and rescaling back to original variables, the amplitude equations:
\begin{subequations}
    \begin{align}
        \dot{B}_{LF} &= \alpha (s - s_{WB}) B_{LF} - \bra{\gamma |B_{LF}|^2 + \eta |B_{RF}|^2 } B_{LF},\\
        \dot{B}_{RF} &= \alpha (s - s_{WB}) B_{RF} - \bra{\gamma |B_{RF}|^2 + \eta |B_{LF}|^2} B_{RF}.
    \end{align}
\end{subequations}

\section{Numerical methods used to obtain nonuniform solutions}\label{app:computational implementation}
\subsection{Continuation and linear stability} \label{sec:cont AUTO}

The travelling waves (TWs), excitable pulses (EPs), travelling fronts (TFs), stationary pulses (SPs), long wavelength oscillatory (LWO), and wave-pinning (WP) solutions discussed in this work are all {steady-state solutions to~\cref{eq:model} in the comoving frame ($\xi=x-ct$) with periodic boundary conditions (or Neumann for TFs). These solutions are computed using continuation with specific additional constraints: (\textit{i}) mass constraint~\cref{eq:mass con} 
\begin{align} \label{eq:mass condition}
\frac{1}{L}\int_0^L[u(\xi)+v(\xi)]{\textrm d}\xi-M=0,
\end{align}
and (\textit{ii}) phase constraint to handle translational (i.e., phase) invariance~\cite{krauskopf_2007,Doedel_2009}
\begin{align} \label{eq:phase condition}
\int_0^L(Q(\xi)-\tilde Q(\xi))^{\rm T}\tilde Q'(\xi){\textrm d}\xi=0,
\end{align}
where $Q=(u,v,F)^{\rm T}$ is the current solution, $\tilde Q$ is the previous solution computed along the branch, and primes denote differentiation with respect to the argument. Consequently, continuation requires three free parameters: the main bifurcation parameter and two additional parameters for the integral constraints. For these constraints, we add two additional ``auxiliary'' parameters, $\kappa$ and $\psi$, which are forced to be small by the integral constraints (i.e., $\kappa,\psi<10^{-12}$ for all solutions). In AUTO, the system must be specified as a first-order ODE system. So, by defining $U:=u'$, $V:=v'$, and $f:=F'$, we rewrite~\cref{eq:model} as:
\begin{align}
    {\begin{aligned} \label{eq:PDE sys TW first}
        u'&=U,\\
        v'&=V,\\
        F'&=f,\\
        DU'&=(1+sF+u^2)u-(b+\gamma u^2)v+\kappa-cU,\\
        V'&=(b+\gamma u^2)v-(1+sF+u^2)u-(c+\psi)V,\\
        D_Ff'&=\theta(F-p_0-p_1u)-cf.
    \end{aligned}}
\end{align}
For the continuation of solutions with $c=0$, we use the free parameters $\{s,\kappa,\psi\}$, and for solutions with $c>0$ the free parameters $\{s,c,\kappa\}$.}
\begin{figure}[tp!]
\centering
    \includegraphics[width=\textwidth]{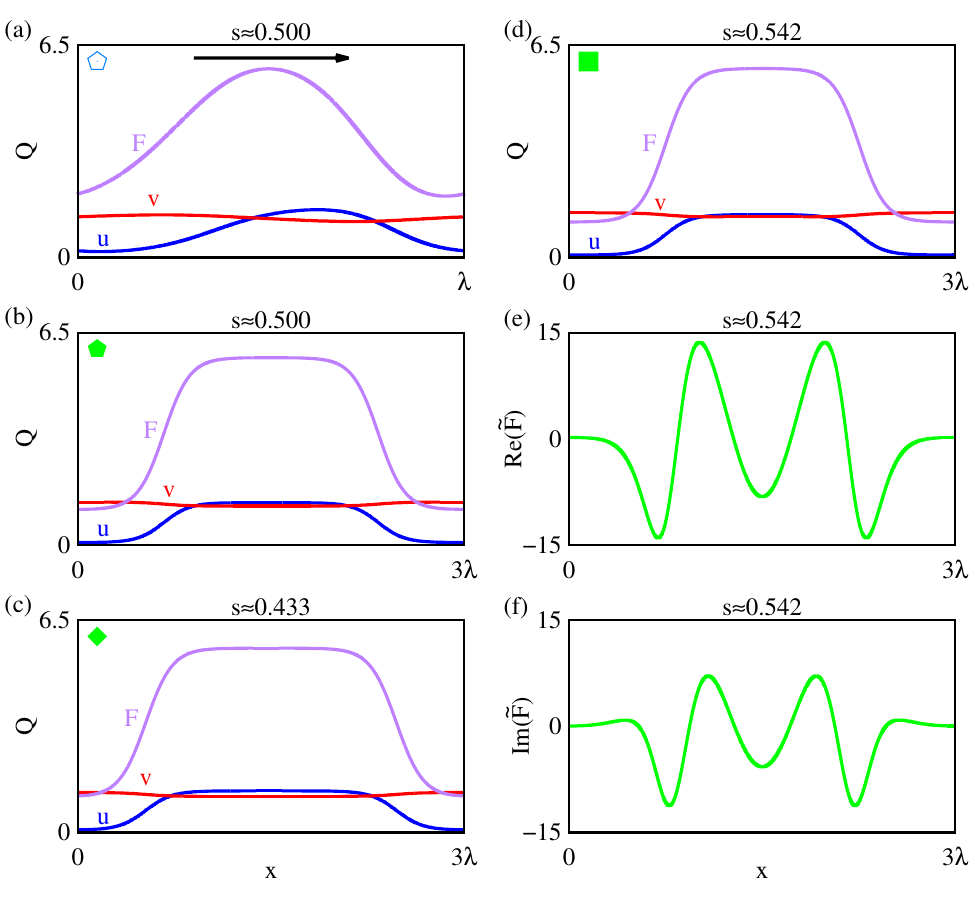}
    \caption{Solution profiles at selected locations in~\cref{fig:sim bifs} showing all solution components for (a) travelling wave (TW) and (b-d) wave-pinning (WP) solutions on periodic domains; the arrow in (a) denotes the direction of the wave. (e) Real and (f) imaginary parts of the $F$-component of the leading eigenfunction ($\widetilde{Q}$) past the Hopf instability of the WP$_{3\lambda}$ branch at $s\approx0.554$ (see~\cref{fig:sim bifs} and~\cref{fig:perturbedWPcell}f where this instability is observed in a time-dependent simulation). The critical complex conjugated eigenvalues at the onset are $\mu\approx \pm 0.648i$. The parameter values are given in~\Cref{tab:par values} with $M=2$ and $b=b_c\approx0.067$.
    } \label{fig:basic sims}
\end{figure}

{Continuation of standing waves (SWs) requires a different approach since such solutions are also time-periodic and are thus two-dimensional (space and time). Since standard continuation in AUTO only deals with one dimension, we use second-order centered finite differencing in space, leaving us with a large system of $3n$ ODEs, where $n$ is the number of discrete points. Since the mass $M$ is independent of time, its temporal average is also $M$, which is used as our mass constraint for SWs.}

{The linear stability of the resulting steady-state solution $Q_{SS}(\xi)$, is determined by investigating perturbations of the form
\begin{align} \label{eq:wave pert}
Q(\xi,t)=Q_{SS}(\xi)+\epsilon \widetilde Q(\xi) e^{\mu t},
\end{align}
leading to the eigenvalue problem
\begin{equation} \label{eq:stab diff operator}
    \mathscr{L} \widetilde Q:=\mathbb{D}\widetilde Q''+c\widetilde Q'+\mathcal{L}(Q_{SS}(\xi))\widetilde Q=\mu \widetilde Q,
\end{equation}
where $\widetilde Q=(\widetilde u,\widetilde v,\widetilde F)^{\rm T}$ denotes the usual components of the model and $\mathbb{D}=\text{diag}(D,1,D_F)$ is a diagonal matrix of the diffusion coefficients. We use second-order centered finite differencing to approximate the linear operator $\mathscr{L}$ in~\cref{eq:stab diff operator}. The solutions from AUTO are converted onto a uniform grid using a built-in splining function in Matlab, $spline$. We compute a subset of the eigenvalues near the origin using the built-in Matlab function $eigs$, ensuring that enough eigenvalues are computed so that the largest real-part eigenvalue is found. To compute the stability of SW solutions, we use the boundary value problem type in AUTO that also performs Floquet multiplier calculations; thus, computing the stability at each step~\cite{Doedel_2009}.}

\subsection{Direct numerical integration} \label{sec:simulation method}
Time-dependent simulations were performed in Julia using the Method of Lines approach~\cite{trefethen_1996}. The spatial domain was discretized using the same second-order centered finite differencing used in the stability analysis (see~\Cref{sec:cont AUTO}). The resulting large system of ODEs was integrated using the $Rodas4P$ method from the Julia DifferentialEquations package~\cite{rackauckas2017differentialequations}, which is a Rosenbrock method purpose-built for semi-discretized non-linear parabolic PDEs~\cite{Steinebach_2023}.

In~\Cref{fig:basic sims}a-d, we show sample profiles of wave-pinning (WP) and travelling wave (TW) solutions used in time-dependent simulations (see~\Cref{sec:time simulations}) showcasing the differences in each component. One wavelength of the TW is used to better distinguish between the differences in solution components. As expected, the solution components are qualitatively similar and all non-negative. In~\Cref{fig:basic sims}e,f, we show the real and imaginary parts of $\widetilde F$ that are associated with the critical eigenfunction $\widetilde Q$ of the Hopf instability of the WP$_{3\lambda}$ branch at $s\approx 0.554$. This demonstrates the oscillatory nature of the instability of WP$_{3\lambda}$ solutions shown in~\cref{fig:perturbedWPcell}f.

\section*{Acknowledgments}
We thank Edgar Knobloch for helpful discussions and comments.

\bibliographystyle{siamplain}
\bibliography{ActinWavesRefs}

\end{document}